\documentclass[preprint,12pt]{aastex}

\newcommand{\beq}{\begin{equation}}           
\newcommand{\eeq}{\end{equation}}             
\newcommand{\bef}{\begin{figure}}           
\newcommand{\enf}{\end{figure}}             

\shorttitle{Limb Darkening Coefficients}

\begin{document}

\title{Computing Limb Darkening Coefficients from Stellar Atmosphere Models}

\author{David Heyrovsk\'y}

\affil{Institute of Theoretical Physics, Charles University\\ V
Hole\v{s}ovi\v{c}k\'ach 2, 18000 Prague, Czech Republic}

\email{david.heyrovsky@mff.cuni.cz}

\begin{abstract}
We explore the sensitivity of limb darkening coefficients computed from stellar atmosphere models to different least-squares fitting methods. We demonstrate that conventional methods are strongly biased to fitting the stellar limb. Our suggested method of fitting by minimizing the radially integrated squared residual yields improved fits with better flux conservation. The differences of the obtained coefficients from commonly used values are observationally significant. We show that the new values are in better agreement with solar limb darkening measurements as well as with coefficients reported from analyses of eclipsing binary light curves.
\end{abstract}


\section{INTRODUCTION}
\label{sec:intro}

With the exception of our Sun, up to recently the observational significance of stellar limb darkening had been limited to the analysis of eclipsing binary light curves \citep{popp84}. A range of new observational techniques and strategies sensitive to limb darkening have been developed since then. These include for example stellar surface interferometry \citep{quirr96}, observation of gravitational microlensing caustic-crossing events \citep{witt95} and extrasolar planet transits \citep{dgc01}. When analyzing any such measurements, expected intensity profiles from stellar atmosphere models are often used in the process of determining other event parameters. However, with sufficiently good quality data these observations can also be used for direct measurement of stellar limb darkening. 

Such measurements are usually done by assuming simple analytical limb darkening laws. The measured coefficients of these laws are then compared with their theoretically predicted values, which are obtained by fitting the laws to model atmosphere intensity profiles. However, these predicted values depend significantly on the adopted method of fitting \citep*[e.g.,][]{dcg95}. Here we demonstrate that even variants of the commonly used least-squares methods can yield very different coefficient values, and suggest a preferred method of fitting. 

In the following \S\ref{sec:methods} we discuss various fitting methods and introduce our suggested method. We study the fit qualities of selected methods in \S\ref{sec:atlas} and demonstrate the differences between obtained coefficients on the example of ATLAS model atmospheres \citep{kur93a}. In \S\ref{sec:sun} we compare these coefficients with measured solar limb darkening, in \S\ref{sec:binaries} with limb darkening coefficient estimates available from eclipsing binary light curve analyses. We discuss variations of the suggested fitting method in \S\ref{sec:discussion} and summarize the main conclusions in \S\ref{sec:summary}.

\section{LIMB DARKENING LAW FITTING METHODS}
\label{sec:methods}

One of the main results obtained from stellar atmosphere models is the
center-to-limb variation of specific intensity $I_\nu$ of the emitted
radiation field computed for a range of wavelengths. The
center-to-limb dependence is obtained by computing radiative transfer along a
set of rays emerging from the atmosphere under different viewing
angles $\theta$. Broadband limb darkening of the intensity $I$
can then be computed from $I_\nu$ for different photometric filter
systems. Intensity values are traditionally parametrized by
$\mu=\cos{\theta}$, ranging from 1 at stellar disk center to 0 at limb, or
alternatively by the radial position on the disk
$r=\sin{\theta}$, ranging from 0 at center to 1 at limb.

Various analytical laws have been suggested for approximating stellar
limb darkening, most of them in the form of linear combinations of simple functions of $\mu$. The most basic limb-darkening law, generally referred to as ``linear limb darkening", is given by
\beq
I_L(\mu)=I_0[1-u(1-\mu)]\,,
\label{eq:lld}
\eeq
where the linear limb-darkening coefficient $u$ determines
the shape of the limb-darkening profile, and $I_0$ is the intensity
of the linear law at disk center. Other laws have been introduced to get a more accurate analytical description, by adding a third term
to equation~(\ref{eq:lld}), for example the quadratic, square-root, or
logarithmic laws \citep[see][]{clar00}. \cite{clar00} also proposed
a more complex law which is a linear combination of five terms.

A range of methods have been used for the seemingly straightforward
task of computing limb-darkening coefficients from the intensity
profiles of model atmospheres, each of them producing generally
different values.

\subsection{PREVIOUS METHODS}
\label{sec:previous}

The two main groups of fitting methods are least-squares based methods (e.g., \citealp*{mbg77,clagi90}; \citealp{dcg95}; \citealp*{cdg95}), and methods requiring flux conservation \citep[e.g.,][]{waru85,vham93}. The primary motivation for the latter methods is the observation that by concentrating on the quality of the fit of the profile shape one often ends up with a low accuracy of flux conservation. This problem is pronounced especially in the case of the linear law, less so in higher order laws. Computation of higher order coefficients in the flux conservation methods requires additional constraints, for example conservation of mean intensity \citep{vham93}, or fixing the intensity value at $\mu=0.1$ close to the limb \citep{waru85}.

Least-squares based methods compute limb-darkening coefficients mostly by minimizing
\beq
\chi^2=\sum_i{[I_L(\mu_i)-I_i]^2}\,,
\label{eq:lsq-discr}
\eeq
where $I_i$ is the intensity of the model atmosphere at disk position $\mu_i$, and $I_L$ is the limb-darkening law of interest. In support of the least-squares methods \cite{clar00} argues that flux-conservation methods fail in the main reason for using analytical laws, which is to get a reasonable fit of the shape of the intensity profile. A good fit to the profile -- particularly in the case of higher order laws -- should inherently lead to sufficiently accurate flux conservation. In comparison, by concentrating on integrated quantities flux-conservation methods largely ignore the shape of the profile.

We note here that in earlier works based on either of the methods (e.g., all those cited in the first paragraph) the central intensity $I_0$ was kept fixed at the model atmosphere value and only the shape-defining limb darkening coefficients were fitted. However, there is no a priori reason to fix one particular point in the intensity profile at its exact value and fit only the remaining points. On the other hand, this approach leads to an overall worse fit. In more recent works this constraint has been apparently dropped \citep{clar00,barb03}. In the case of the linear law from equation~(\ref{eq:lld}) the simplest approach is to fit the linear parameters $I_0$ and $u\,I_0$ and compute the limb darkening coefficient $u$ as the ratio of the obtained values. All the other laws mentioned above can be fitted in a similar manner.

The least-squares method has its own further ambiguities. As pointed
out by \cite{dcg95}, the obtained coefficients depend on the
distribution of the rays $\mu_i$ provided by the model atmospheres.
For example, in the commonly used Kurucz ATLAS9 models \citep{kur93a,kur93b,kur93c,kur94}
the spacing of the 17 computed rays decreases sharply at low values
of $\mu$, close to the limb. A straightforward fit minimizing
equation~(\ref{eq:lsq-discr}) then gives a high weight to the
quality of the fit close to the limb on account of the fit quality
closer to the center. In order to get a less biased fit \cite{dcg95}
as well as subsequently \cite{cdg95} or \cite{clar00} used only a
subset of the available rays more evenly spaced in $\mu$. Similarly,
\cite{barb03} who used their own ATLAS-based model atmospheres
computed 20 rays evenly distributed in $\mu$.

\subsection{SUGGESTED METHOD}
\label{sec:suggested}

A simple way to overcome problems caused by the spacing of discrete points $[\mu_i,I_i]$ is to suitably interpolate the intensity between the points, so that we get a continuous function $\tilde{I}(\mu)$ which passes through all the model points, $\tilde{I}(\mu_i)=I_i$. For this purpose we can use cubic spline interpolation. The $I_i$ versus $\mu_i$ dependence is typically smooth, without large jumps or wiggles, so that such a spline gives a reasonable interpolation. The limb darkening coefficients can now be computed by minimizing the integrated squared residual
\beq
\Delta^2_{\mu}=\int[I_L-\tilde{I}\,]^2\,d\mu ,
\label{eq:lsq-mu1}
\eeq
where all integrals appearing in the least-squares solution can be evaluated to arbitrary precision. The result obtained in this way is what previous authors were trying to achieve by selecting regularly spaced points. In the suggested approach none of the points obtained from the model atmosphere computations are omitted, thus no useful information is neglected. The spline interpolation provides a sensible guess about the behavior of the intensity in between these points. We discuss the sensitivity of the results to the particular choice of spline interpolation in \S\ref{sec:discussion}.

However, the ambiguity due to the spacing of points is overcome only
seemingly. The new ambiguity lies in the choice of the integration
variable in equation~(\ref{eq:lsq-mu1}). In this example,
integration over $\mu$ gives equal weight to the squared residual in
equal-sized intervals of $\mu$. For instance, the squared residual
in the radial interval from the center to $r\doteq 0.87$
(corresponding to $\mu=0.5$) has the same weight as in the remaining
interval close to the limb. By transforming from $\mu$ to the radial
coordinate $r$ in the integral in equation~(\ref{eq:lsq-mu1}) we get
\beq
\Delta^2_{\mu}=\int[I_L-\tilde{I}\,]^2\,r(1-r^2)^{-1/2}d r .
\label{eq:lsq-mu2}
\eeq
We see that the residual at the disk center has zero weight and the weight function grows monotonously towards the limb where it diverges. The choice of $\mu$ as the integration variable thus still gives results strongly biased to fitting the limb of the star. In order to get a reasonable fit even closer to the center, we decide to integrate over $r$ and compute limb darkening coefficients by minimizing
\beq
\Delta^2_{r}=\int[I_L-\tilde{I}\,]^2\,d r .
\label{eq:lsq-r}
\eeq

The suitability of the choice of integration variable (or equivalently weighting) should be primarily demonstrated in confrontation with observational data, and we do so in \S\ref{sec:observations}. In support of our choice we briefly consider here some relevant astrophysical situations, namely, eclipsing binaries and caustic-crossing microlensing events. In either of these cases the weighting kernel that is convolved with the intensity to obtain the flux is time-dependent, depending on the phase of the binary or the position of the passing lens, respectively. Hence neither scenario directly provides a single ideal weighting. However, if we look for simplicity at the case of a $90^\circ$ inclination binary, the radial position of the occulting component shifts linearly with time. Moreover, for example in the case of annular eclipses the occulted area is of constant radial size. Similarly, in zero impact parameter single-lens microlensing the radial position of the lens changes linearly with time, at each position giving maximum weight to the point directly behind the lens. Such considerations suggest that equal radial weighting is a natural and reasonable choice.

In the following section we illustrate the differences between the results obtained by the discussed fitting methods on the example of widely used model atmospheres.

\section{DEMONSTRATION ON ATLAS ATMOSPHERES}
\label{sec:atlas}

We demonstrate the different fitting methods on the ATLAS9 model atmospheres computed by \cite{kur93a,kur93b,kur93c,kur94} using the intensity files available on Robert Kurucz's website\footnote{\url{http://kurucz.harvard.edu}}. The models in the parameter grid have metallicities ranging from $[Fe/H]=+1$ to -5, effective temperatures from $T_{eff}= 3500$ to $50000\,K$, surface gravities in CGS units from $log\,g= 0$ to 5, and microturbulent velocities for solar metallicity models from $v_t=0$ to 8 $km\, s^{-1}$, for other metallicities only 2 $km\, s^{-1}$. For each model specific intensity $I_\nu$ is given for wavelengths from $\lambda=9.09\,nm$ to $160\,\mu m$ spaced by $2\,nm$ in the optical range, and for 17 rays passing through the atmosphere: 10 at $\mu$ values from 1 to 0.1 spaced by 0.1, and 7 additional rays close to the limb at 0.25, 0.15, 0.125, 0.075, 0.05, 0.025, 0.01 (see Figure~\ref{fig:fits}). We obtain intensity profiles for specific photometric filters by integrating
\beq
I_i(\mu_i)=\int\frac{c}{\lambda^2}I_\nu(\mu_i,\lambda)\,S(\lambda)\,
d\lambda\,,
\label{eq:intensity}
\eeq
where $c/\lambda^2$ is the specific intensity conversion factor (from per frequency to per wavelength), and $S(\lambda)$ is the filter response function. In this work we use the standard $BV\!RI$ filter response functions given by \cite{bess90}. We formally extend the intensity profile to $\mu=0$ by linear extrapolation from the two points closest to the limb, and obtain $\tilde{I}(\mu)$ by interpolating the profile using cubic splines with natural boundary conditions. We discuss the influence of different boundary conditions in \S\ref{sec:discussion}.

We compare four different least-squares fitting methods. First, we fit all 17 points provided by Kurucz's models minimizing $\chi^2$ given by equation~(\ref{eq:lsq-discr}). Second, we use the same equation to fit a subset of 11 points with 10 regularly spaced $\mu$ values from 1 to 0.1 plus $\mu=0.05$ -- these are the points fitted by \cite{dcg95}, \cite{cdg95} or \cite{clar00}. Third, we fit the spline-interpolated intensity $\tilde{I}$ minimizing $\Delta^2_{\mu}$ from equation~(\ref{eq:lsq-mu1}) using integration over $\mu$. Fourth is our suggested method of fitting $\tilde{I}$ by minimizing $\Delta^2_r$ from equation~(\ref{eq:lsq-r}) using integration over $r$.

In Figure~\ref{fig:fits} we present as an example the V band profile of a $T_{eff}=4250\,K$, $log\,g=4.5$, $[Fe/H]=-3.5$ model atmosphere plotted separately as a function of $r$ and $\mu$. The plots include the original points and their spline-interpolated intensity profile $\tilde{I}$ (here normalized to unit $r-$integrated rms intensity), as well as linear limb darkening laws fitted by the four methods and their residuals from $\tilde{I}$. As expected and as pointed out by \cite{dcg95}, the 17-point fit has the heaviest bias to the limb. The 11-point fit here closely approximates the $mu-$integrated fit which has the lowest $mu$-integrated squared residual (see lower right panel), and both are only slightly less biased to the limb. The $r-$integrated fit by definition has the lowest $r-$integrated squared residual (see lower left panel). The non-integrated residuals of this fit are overall the smallest from the center to roughly $\mu=0.2$ (i.e., $r=0.98$). Although the quality of the $\mu-$integrated fit is substantially better on the remaining interval, the radial width of this interval is only 0.02. We conclude that the $r-$integrated fit gives the best linear limb darkening approximation of the intensity profile.

In order to compare the fitting methods more generally, we fitted $BV\!RI$ profiles of the full range of models described above, a set comprising altogether 38324 profiles of 9581 model atmospheres. Besides the linear law we used quadratic, square-root, logarithmic, and Claret limb-darkening laws \citep[for definitions see][]{clar00}. For each fit we computed two parameters as measures of fit quality. The first is the relative rms residual defined as the ratio of rms residual to rms intensity
\beq
\sigma=\frac{\Delta_r}{\sqrt{\int\tilde{I}^2dr}}\,.
\label{eq:res}
\eeq
Our second parameter is the relative flux excess of the obtained fit over the original profile
\beq
\frac{\Delta F}{F}=\frac{\int I_L r\,dr}{\int\tilde{I}r\,dr}\,-1\,.
\label{eq:fluxdev}
\eeq
The average and maximum absolute values of these parameters for the five limb darkening laws and four fitting methods are presented in Table~\ref{tab:compare}. Judging by the residuals $\sigma$, for all the laws $r-$integration by definition gives the best results (for the linear law average $\sigma=1.25\%$, worst fitted profile has $\sigma=3.53\%$), followed by the substantially worse 11-point fit (2.34\%, 5.55\%), $\mu-$integration (2.71\%, 6.28\%), and the poorest 17-point fit (3.33\%, 6.99\%).

Comparing the fitting methods by relative flux excess yields a different picture. For all the laws $\mu-$integration gives by far the best flux conservation (here limited by machine precision, for the linear law formally average $|\Delta F/F|=1.18\times10^{-16}$, the highest value for a profile $|\Delta F/F|=5.96\times10^{-16}$), followed by $r-$integration ($2.27\times10^{-3}$, $8.65\times10^{-3}$). Except in the case of the linear law where the order of the remaining methods is reversed, these are followed by the 11-point fit ($4.38\times10^{-3}$, $1.29\times10^{-2}$) and the 17-point fit ($3.63\times10^{-3}$, $1.42\times10^{-2}$). The reason for the excellent flux conservation\footnote{ Note that accuracy better than $\sim 10^{-4}$ exceeds the flux accuracy of the underlying model atmosphere data, as shown in \S\ref{sec:discussion}.} in $\mu-$integration is the similar weighting of the intensity in the flux integral, as seen in equation~(\ref{eq:fluxdev}). The factor $r$ gives zero weight at the disk center and highest (though not divergent) weight at the limb. Nevertheless, in view of the roughly two times higher average residual in comparison with $r-$integration for the four simpler limb darkening laws, we conclude that this method suffers from a similar problem as the non-least squares flux-conservation methods \citep{waru85,vham93}, i.e., achieving good flux conservation at the cost of a poor fit of the intensity profile shape. The obtained results support the use of $r-$integration fitting of the spline-interpolated intensity profile as the method of choice for computing limb darkening coefficients.

We note that Table~\ref{tab:compare} can be also used for example to compare the overall quality of the limb darkening laws for the different fitting methods. However, discussions of the different analytical laws are beyond the scope of this paper.

The main result of this section for practical purposes is the numerical difference between the limb darkening coefficients produced by different fitting methods. We limit ourselves here to comparing those obtained by our suggested method of $r-$integration (hereafter $u_r$) with those obtained by the 11-point fit (hereafter $u_{11}$), the most frequently used coefficients \citep[e.g.,][]{dcg95,cdg95,clar00}. For illustration we plot $BV\!RI$ values of $u_r$ in Figure~\ref{fig:coeff-solarcomp} as a function of $T_{eff}$. In order to avoid crowding we include only solar abundance models with $v_t=2\,km\,s^{-1}$. The specific range of $log\,g$ of the models depends on $T_{eff}$. The main tendencies are well known: overall the coefficient decreases with increasing temperature and increasing wavelength. The rms residuals and flux excesses for these models and both types of fitting are shown in Figure~\ref{fig:resid-solarcomp} as a function of $T_{eff}$. The $r-$integrated fits have lower residuals and better flux conservation, in agreement with the results in Table~\ref{tab:compare}.

Figure~\ref{fig:allcoeffdiff} illustrates the difference between $u_r$ and $u_{11}$ for the full range of model atmospheres, plotted as a function of $u_{11}$. The coefficient difference is substantial, ranging from -0.14 to 0.11 with most of the $u_r$ lower than $u_{11}$, corresponding to flatter limb darkening profiles. At the same time, the most limb darkened profiles are even more limb darkened when using $u_r$. The results are plotted again in Figure~\ref{fig:bandcoeffdiff} as a function of $T_{eff}$ for each photometric band separately. The greatest variation occurs for lower temperature models ($T_{eff}\lesssim7000\,K$) and is much reduced for higher temperature models ($T_{eff}\gtrsim10000\,K$). The range of coefficient differences decreases from the $B$ band to the $I$ band. We note specifically that 11-point fitting overestimates the limb darkening coefficient in the $I$ band for all computed models, in the $R$ band for all models with $T_{eff}>4500\,K$, in the $V$ band for all models with $T_{eff}>6000\,K$, and in the $B$ band for all models with $T_{eff}>7000\,K$. At lower temperatures coefficients are overestimated for some profiles and underestimated for others.

A comparison of individual coefficients of higher order limb darkening laws reveals there can be even greater differences between the values obtained by different fitting methods. Nevertheless, as seen from Table~\ref{tab:compare}, at the same time the overall fit quality is better, thus the obtained combinations of coefficients produce more similar limb darkening shapes.

\section{COMPARISON WITH OBSERVATIONAL RESULTS}
\label{sec:observations}

Following the discussions of relative merits of different limb darkening law fitting methods we turn to the question of observational relevance. We note that the differences between the linear coefficients demonstrated in the previous section are large enough to be detectable not only in solar limb darkening measurements, but for example also in observations of caustic-crossing microlensing events or eclipsing binaries. In \cite{cass06} we showed that the coefficients measured in caustic-crossing microlensing events involving K giants in the Galactic Bulge are in better agreement with the values obtained by $r-$integration than with the 11-point fit values of \cite{clar00}. Here we present a comparison of these coefficients using solar limb darkening measurements and eclipsing binary results.

\subsection{SOLAR LIMB DARKENING}
\label{sec:sun}

For our comparison we use measured solar limb darkening data from \citet[][p.356]{cox00} for the four best-sampled wavelengths in the optical range: 450, 500, 550, and $600\,nm$. We compare these limb-darkening profiles with linear limb darkening fits for the same wavelengths of the \cite{kur94} solar model with $T_{eff}=5777\,K$, $log\,g=4.4377$, $[Fe/H]=0.0$, $v_t=1.5\,km\, s^{-1}.$ The $r-$integrated and 11-point fits give us coefficients $u_r$ and $u_{11}$, respectively. We take these fixed coefficients and fit only the central intensity $I_0$ from equation~(\ref{eq:lld}) to get a best fit to the spline-interpolated solar limb darkening.

The relative residuals of these fits are plotted in Figure~\ref{fig:sun} for each of the wavelengths as a function of $r$. The values of the relative rms residual $\sigma$ and relative flux excess $\Delta F/F$ are given in each of the panels. We see that for all four wavelengths $u_r$ gives better results than $u_{11}$, its $\sigma$ is lower by a factor of 1.3 -- 1.7 and $\Delta F/F$ is lower by a factor of 5 -- 21. In all cases $u_{11}$ yields a much steeper limb darkening profile than $u_r$ and the measured profile. We conclude that the theoretical limb darkening coefficient obtained by $r-$integrated fitting of the model atmosphere data is in substantially better agreement with the measured solar limb darkening than the coefficient obtained by 11-point fitting.

\subsection{LIMB DARKENING FROM ECLIPSING BINARY ANALYSES}
\label{sec:binaries}

Obtaining limb darkening coefficients from the analysis of eclipsing binary light curves is no simple task. Already \cite{popp84} estimated that getting a limb darkening coefficient with a precision of 0.1 or better requires at least 100 light curve points within minima with a 0.005 mag photometric accuracy. Not surprisingly, there are only a few eclipsing binary systems for which limb darkening coefficients have been extracted and published. These limb darkening measurements have been obtained by different teams using different methods, approximations, and approaches -- and are therefore also weighted by different problems. In this section we do not re-analyze any of the binary light curves. We merely perform the simplest test for our purposes, collating the published values and checking their agreement with theoretical coefficients obtained from model atmospheres as described above.

More specifically, we searched the literature for eclipsing binaries with measured limb darkening coefficients, for which estimates of $T_{eff}$, $log\,g$, and $[Fe/H]$ were also available (we note here that these estimates may sometimes also be problematic). In a number of these binaries the coefficients for the primary and secondary stars were not measured independently. Instead, their difference was kept fixed at a value expected from some specific model atmosphere coefficients. We restrict our sample to the systems analyzed without this constraint plus to those for which the fixed difference values agree within 0.01 with those expected from our $u_r$ coefficients \emph{as well as} \cite{clar00} coefficients (hereafter $u_C$). The measured $BV\!R$ linear limb darkening coefficients for these nine binaries are listed in Table~\ref{tab:eclbin} together with the corresponding values of $u_C$ and $u_r$. These were obtained by trilinear interpolation of the values from \cite{clar00}, and of the values computed for the Kurucz ATLAS9 model atmosphere grid \citep{kur93a,kur93b,kur93c} by $r-$integration fitting, respectively, using the observed stars' $T_{eff}$, $log\,g$, and $[Fe/H]$ values.

For five of the binaries (BS Dra, WW Cam, V459 Cas, WW Aur, RW Lac) all measured coefficients are closer to the $u_r$ values, for MU Cas they are closer to the $u_C$ values, while for the remaining three (EE Peg, FS Mon, GG Ori) some are closer to $u_r$ and others to $u_C$. Looking at the latter four systems, in the case of MU Cas we note that \cite*{lcs04b} first obtained $u=0.32$ for both components in an unconstrained fit -- a value in perfect agreement with $u_r$. The value quoted in the table was obtained by fitting keeping a fixed luminosity ratio. Turning to EE Peg \citep{lapo84} we see that $u_r$ values for both components are also in agreement with the measurements. The FS Mon coefficients \citep{lacy00} have no error bars\footnote{This is also the case with RW Lac.} and the agreement with $u_r$ in the $V$ band is only marginally worse than with $u_C$, while the agreement with $u_C$ in the $B$ band is substantially worse than with $u_r$. In the last system, GG Ori \citep{torr00}, the $B$ and $R$ values agree better with $u_r$ while the $V$ coefficients agree better with $u_C$. Here the situation is similar to MU Cas: the original unconstrained fits from two datasets produced $V$ band coefficients $0.48\pm0.08$ and $0.41\pm0.05$, values better consistent with $u_r=0.45$ than with $u_C=0.51$. The unconstrained $R$ band value of $0.33\pm0.13$ is also in much better agreement with $u_r=0.38$ than the value quoted in the table.

The limb darkening of eclipsing binaries in which the radii of the components are a sufficiently large fraction of the orbital radius ($>$ 10 - 15\%) may be influenced by the reflection effect. In such systems the measured limb darkening coefficient may not agree with the value given by the star's $T_{eff}$. In order to avoid this potential effect, we check the results for binaries in our selection with the smallest radii. These are RW Lac, V459 Cas, and GG Ori, with radii in the range 4 - 7\%. From the values in Table~\ref{tab:eclbin} and the discussion of GG Ori above we see that the conclusions about the comparison of the two coefficients remain valid.

Just as in the case of caustic-crossing microlensing events, our conclusion is that for eclipsing binaries limb darkening coefficients obtained by our suggested method give overall better results than the widely used \cite{clar00} coefficients. The systems in which original unconstrained fits give better agreement with expected values than photometrically adjusted fits indicate possible problems caused by poor flux conservation or more generally the inadequacy of the linear limb darkening law.

\section{DISCUSSION}
\label{sec:discussion}

When comparing the fitting methods in \S\ref{sec:atlas} one could also evaluate the relative $\mu-$integrated rms residual, differing from $\sigma$ in equation~(\ref{eq:res}) by using $\Delta_{\mu}$ in the numerator and integrating over $\mu$ instead of $r$ in the denominator. Such a residual is by definition lowest for the results obtained by $\mu-$integration. For the linear law we get an average residual $3.28\%$ and maximum residual $6.94\%$ -- both values higher than the corresponding $r-$integrated residuals. We get similar results even for the second order laws. This only demonstrates that fitting of simple analytical laws is more difficult in terms of $\mu$ than $r$. As noted before, fitting by integration over $\mu$ gives very high weight to the limb, where attempts to fit the more complex intensity dependence go on account of the quality of the remaining fit. Only Claret's 5-term law has an average $\mu-$integrated residual lower than its average $r-$integrated residual, even though its maximum $\mu-$integrated residual remains higher. This law therefore tends to fit the limb darkening close to the limb slightly better than closer to the center.

In this work we interpolated intensity data points using cubic spline interpolation in terms of $\mu$ with natural boundary conditions, i.e., zero second derivative at $\mu=0$ and $\mu=1$. In order to check the sensitivity of the results to the interpolation method, we tested using cubic spline interpolation in terms of $r$ with zero first derivative at center and zero second derivative at limb. We note that this interpolation is not esthetically optimal, as in the region closest to the limb the $r$ dependence is very steep, which can produce small wiggles in the spline between the last points, noticeable when plotted as a function of $\mu$. Still, the obtained linear coefficients differ at most in the third decimal place.

Similarly we tested the influence of different boundary conditions for the $\mu$ spline. Changing the boundary condition at the limb or the method of extrapolating the limb intensity had a negligible influence, on the order of $10^{-7}$ in flux or less. Changing the boundary condition at the center to equality of first and second derivatives (which gives an interpolation equally pleasing to the eye) had a larger influence, though still only on the order of $10^{-4}$ in flux or less. The difference is due to the larger spacing of model atmosphere rays at the center. Based on this test we can consider the value of $10^{-4}$ as a limiting flux accuracy for the model atmosphere data used in this work.

\section{SUMMARY AND CONCLUSIONS}
\label{sec:summary}

We have studied the sensitivity of limb darkening coefficients computed from stellar atmosphere models to different least-squares fitting methods. Tests with Kurucz ATLAS9 model atmospheres \citep{kur93a,kur93b,kur93c,kur94} show that best results are obtained by suitably interpolating the available intensity points and fitting limb-darkening laws to the obtained intensity profile $\tilde{I}$ by minimizing the $r-$integrated squared residual introduced in equation~(\ref{eq:lsq-r}). Other methods give too much weight to the quality of the fit in a narrow region at the stellar limb.

The differences between the coefficient values computed by the different methods are substantial. In particular, we have shown that linear limb darkening coefficients of $BV\!RI$ profiles of Kurucz models computed by the suggested method differ by as much as 0.14 from values obtained by a commonly used fitting method with points roughly evenly spaced in terms of $\mu$. Such differences are observationally significant. The plots in Figure~\ref{fig:sun} illustrate that coefficients computed from a solar model atmosphere by the preferred fitting are in better agreement with solar limb darkening data. In Table~\ref{tab:eclbin} we have further demonstrated that available limb darkening measurements from eclipsing binaries are more compatible with the newly computed coefficients than with those from \cite{clar00}. We have reached a similar conclusion previously with limb darkening measurements from caustic-crossing gravitational microlensing events \citep{cass06}.

Other methods of measuring limb darkening such as stellar interferometry or observation of extrasolar planet transits can use a similar approach when comparing measured coefficients with theoretical predictions. However, all these methods may run into problems with analytical limb darkening laws. The limb darkening signature in the data is usually not strong enough for a reliable extraction of more than a single limb darkening coefficient, i.e., the linear coefficient. At the same time, even at the photometric accuracy of the available data the linear law is in many cases an inadequate approximation, largely due to its poor flux conservation and generally poor matching of the limb darkening shape.

The drawbacks of measuring coefficients of analytical laws can be bypassed by using an alternative description of limb darkening, obtained by principal component analysis of an ensemble of model atmosphere intensity profiles \citep{hey03}. At the cost of not being analytical, such an approach gives the best possible limb darkening description as a linear combination of the least number of terms. The principal component description is therefore also ideally suited for measuring stellar limb darkening, as demonstrated by \cite{hey03} in the case of microlensing events.

\acknowledgements

It is a pleasure to thank Jean-Philippe Beaulieu for his hospitality at the Institut d'Astrophysique de Paris. This work was supported by the Czech Science Foundation grant GACR 205/04/P256.

\clearpage
\begin{deluxetable}{llcccc}

\tablecaption{Limb darkening fit quality for different least-squares fitting methods\label{tab:compare}}
\tablehead{\multicolumn{1}{l}{LD law} & \multicolumn{1}{l}{Method} & \colhead{Average $\sigma$} & \colhead{Max. $\sigma$} & \colhead{Average $|\Delta F/F|$} & \colhead{Max. $|\Delta F/F|$} \\ & & (\%) & (\%) & & }

\startdata

Linear & 17-point & 3.327 & 6.987 & $3.63\times 10^{-3}$ & $1.42\times 10^{-2}$ \\

& 11-point & 2.344 & 5.554 & $4.38\times 10^{-3}$ & $1.29\times 10^{-2}$ \\

& $\mu-$int. spline & 2.706 & 6.275 & $1.18\times 10^{-16}$ & $5.96\times 10^{-16}$ \\

& $r-$int. spline & 1.253 & 3.527 & $2.27\times 10^{-3}$ & $8.65\times 10^{-3}$ \\

\tableline

Quadratic & 17-point & 1.180 & 3.303 & $3.85\times 10^{-3}$ & $1.07\times 10^{-2}$ \\

& 11-point & 0.653 & 1.785 & $1.10\times 10^{-3}$ & $2.98\times 10^{-3}$ \\

& $\mu-$int. spline & 0.980 & 2.609 & $6.38\times 10^{-16}$ & $1.75\times 10^{-15}$ \\

& $r-$int. spline & 0.417 & 0.991 & $2.35\times 10^{-4}$ & $6.15\times 10^{-4}$ \\

\tableline

Square-root & 17-point & 0.498 & 3.545 & $1.13\times 10^{-3}$ & $1.08\times 10^{-2}$ \\

& 11-point & 0.288 & 2.115 & $3.32\times 10^{-4}$ & $4.71\times 10^{-3}$ \\

& $\mu-$int. spline & 0.415 & 3.256 & $4.21\times 10^{-16}$ & $1.65\times 10^{-15}$ \\

& $r-$int. spline & 0.238 & 1.382 & $1.28\times 10^{-4}$ & $1.99\times 10^{-3}$ \\

\tableline

Logarithmic & 17-point & 0.665 & 2.657 & $1.85\times 10^{-3}$ & $8.70\times 10^{-3}$ \\

& 11-point & 0.345 & 1.587 & $5.35\times 10^{-4}$ & $3.41\times 10^{-3}$ \\

& $\mu-$int. spline & 0.528 & 2.363 & $1.85\times 10^{-16}$ & $7.91\times 10^{-16}$ \\

& $r-$int. spline & 0.259 & 0.999 & $1.66\times 10^{-4}$ & $1.18\times 10^{-3}$ \\

\tableline

Claret & 17-point & 0.190 & 0.625 & $8.27\times 10^{-5}$ & $5.05\times 10^{-4}$ \\

& 11-point & 0.175 & 0.558 & $4.57\times 10^{-5}$ & $2.17\times 10^{-4}$ \\

& $\mu-$int. spline & 0.184 & 0.585 & $4.48\times 10^{-16}$ & $2.62\times 10^{-15}$ \\

& $r-$int. spline & 0.170 & 0.550 & $1.81\times 10^{-5}$ & $5.48\times 10^{-5}$ \\

\enddata

\tablecomments{Average and maximum values of relative rms residual $\sigma$ and relative flux excess $|\Delta F/F|$ evaluated for $BV\!RI$ profiles of full range of Kurucz model atmospheres (see text).}

\end{deluxetable}

\clearpage
\begin{deluxetable}{lcccccccccccl}

\tabletypesize{\scriptsize}
\tablecaption{Linear limb darkening coefficients measured in eclipsing binaries\label{tab:eclbin}}
\tablehead{\multicolumn{2}{l}{Binary} & \multicolumn{3}{c}{Measured $u$} & \multicolumn{3}{c}{Model $u_C$} & \multicolumn{3}{c}{Model $u_r$} & \colhead{Ref.}\\ & & \colhead{$B$} & \colhead{$V$} & \colhead{$R$} & \colhead{$B$} & \colhead{$V$} & \colhead{$R$} & \colhead{$B$} & \colhead{$V$} & \colhead{$R$} &}

\startdata

BS Dra & A & $0.64\pm0.04$ & $0.50\pm0.03$ & \ldots & 0.707 & 0.605 & \ldots & \bfseries 0.651 & \bfseries 0.539 & \ldots & 1 \\
& B & $0.64\pm0.04$ & $0.50\pm0.03$ & \ldots & 0.706 & 0.605 & \ldots & \bfseries 0.650 & \bfseries 0.539 & \ldots & \\

EE Peg & A & $0.62\pm0.03$ & \ldots & \ldots & 0.660 & \ldots & \ldots & \bfseries 0.597 & \ldots & \ldots & 2 \\
& B & $0.75\pm0.15$ & \ldots & \ldots & \bfseries 0.740 & \ldots & \ldots & 0.689 & \ldots & \ldots & \\

FS Mon & A & 0.58 & 0.58 & \ldots & 0.710 & \bfseries 0.607 & \ldots & \bfseries 0.654 & 0.540 & \ldots & 3 \\
& B & 0.594 & 0.591 & \ldots & 0.718 & \bfseries 0.614 & \ldots & \bfseries 0.664 & 0.548 & \ldots & \\

GG Ori & A & $0.50\pm0.04$ & $0.51\pm0.03$ & $0.23\pm0.07$ & 0.594 & \bfseries 0.513 & 0.432 & \bfseries 0.521 & 0.446 & \bfseries 0.377 & 4 \\
& B & $0.50\pm0.04$ & $0.51\pm0.03$ & $0.23\pm0.07$ & 0.594 & \bfseries 0.513 & 0.432 & \bfseries 0.521 & 0.446 & \bfseries 0.377 & \\

WW Cam & A & \ldots & $0.494\pm0.017$ & \ldots & \ldots & 0.601 & \ldots & \ldots & \bfseries 0.533 & \ldots & 5 \\
& B & \ldots & $0.499\pm0.017$ & \ldots & \ldots & 0.606 & \ldots & \ldots & \bfseries 0.538 & \ldots & \\

V459 Cas & A & \ldots & $0.487\pm0.008$ & \ldots & \ldots & 0.546 & \ldots & \ldots & \bfseries 0.479 & \ldots & 6 \\
& B & \ldots & $0.487\pm0.008$ & \ldots & \ldots & 0.548 & \ldots & \ldots & \bfseries 0.481 & \ldots & \\

MU Cas & A & \ldots & $0.56\pm0.07$ & \ldots & \ldots & \bfseries 0.392 & \ldots & \ldots & 0.331 & \ldots & 7 \\
& B & \ldots & $0.56\pm0.07$ & \ldots & \ldots & \bfseries 0.399 & \ldots & \ldots & 0.337 & \ldots & \\

WW Aur & A & $0.616\pm0.056$ & $0.416\pm0.060$ & \ldots & 0.692 & 0.598 & \ldots & \bfseries 0.625 & \bfseries 0.528 & \ldots & 8 \\
& B & $0.512\pm0.078$ & $0.418\pm0.083$ & \ldots & 0.688 & 0.593 & \ldots & \bfseries 0.618 & \bfseries 0.518 & \ldots & \\

RW Lac & A & \ldots & 0.55 & \ldots & \ldots & 0.668 & \ldots & \ldots & \bfseries 0.611 & \ldots & 9 \\
& B & \ldots & 0.57 & \ldots & \ldots & 0.689 & \ldots & \ldots & \bfseries 0.638 & \ldots & \\

\enddata

\tablecomments{For each eclipsing binary primary star marked by A, secondary by B. Model coefficients based on Kurucz ATLAS9 atmospheres: $u_C$ taken from \cite{clar00}, $u_r$ computed in this work. Better agreeing coefficient is marked in boldface.}
\tablerefs{(1) \citealt{popetz81}; (2) \citealt{lapo84}; (3) \citealt{lacy00}; (4) \citealt{torr00}; (5) \citealt{lacy02}; (6) \citealt*{lcs04a}; (7) \citealt{lcs04b}; (8) \citealt{south05}; (9) \citealt{lacy05}}

\end{deluxetable}

\clearpage
\bef[t]
\begin{center}
\includegraphics[scale=.4]{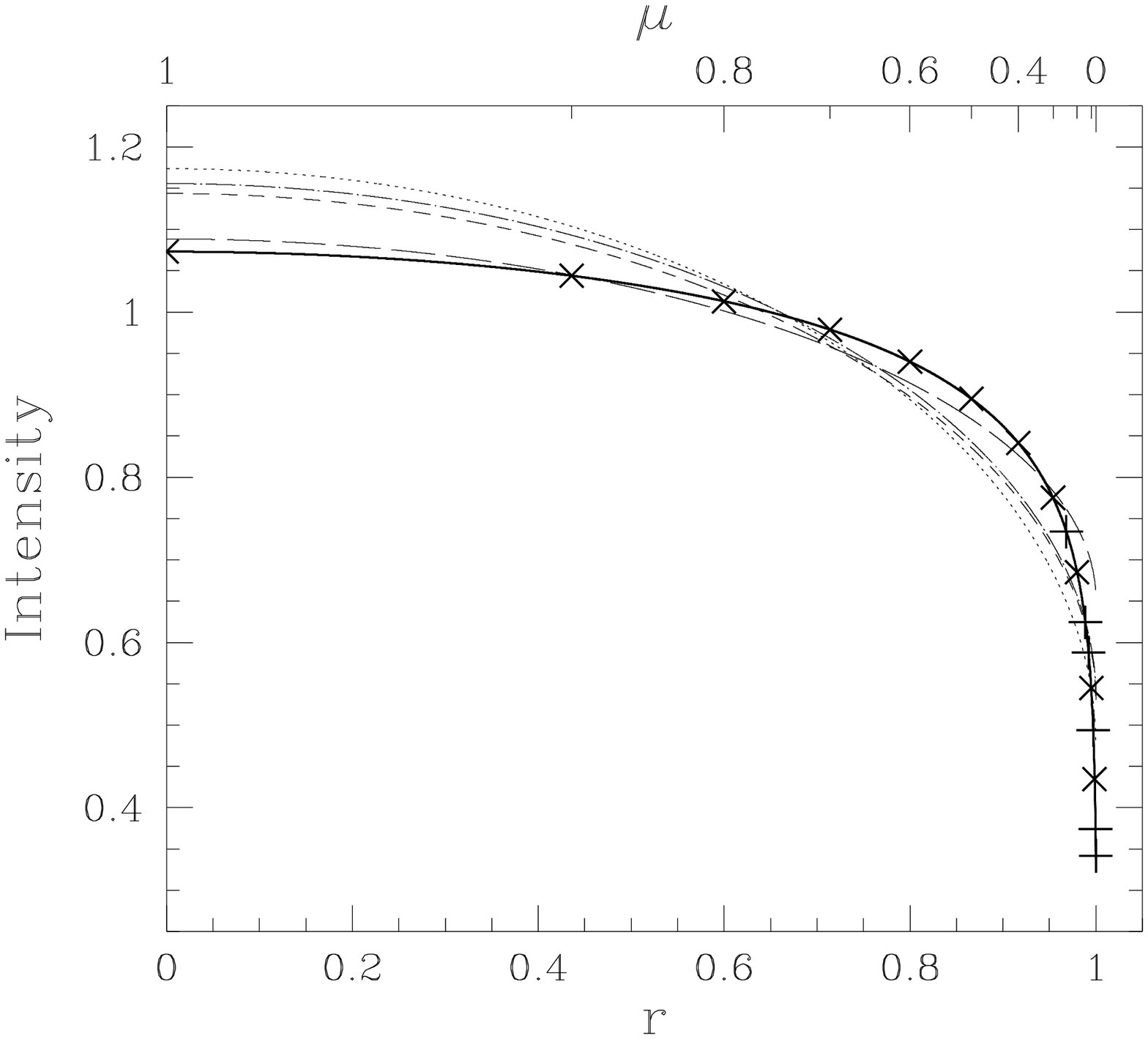}
\includegraphics[scale=.4]{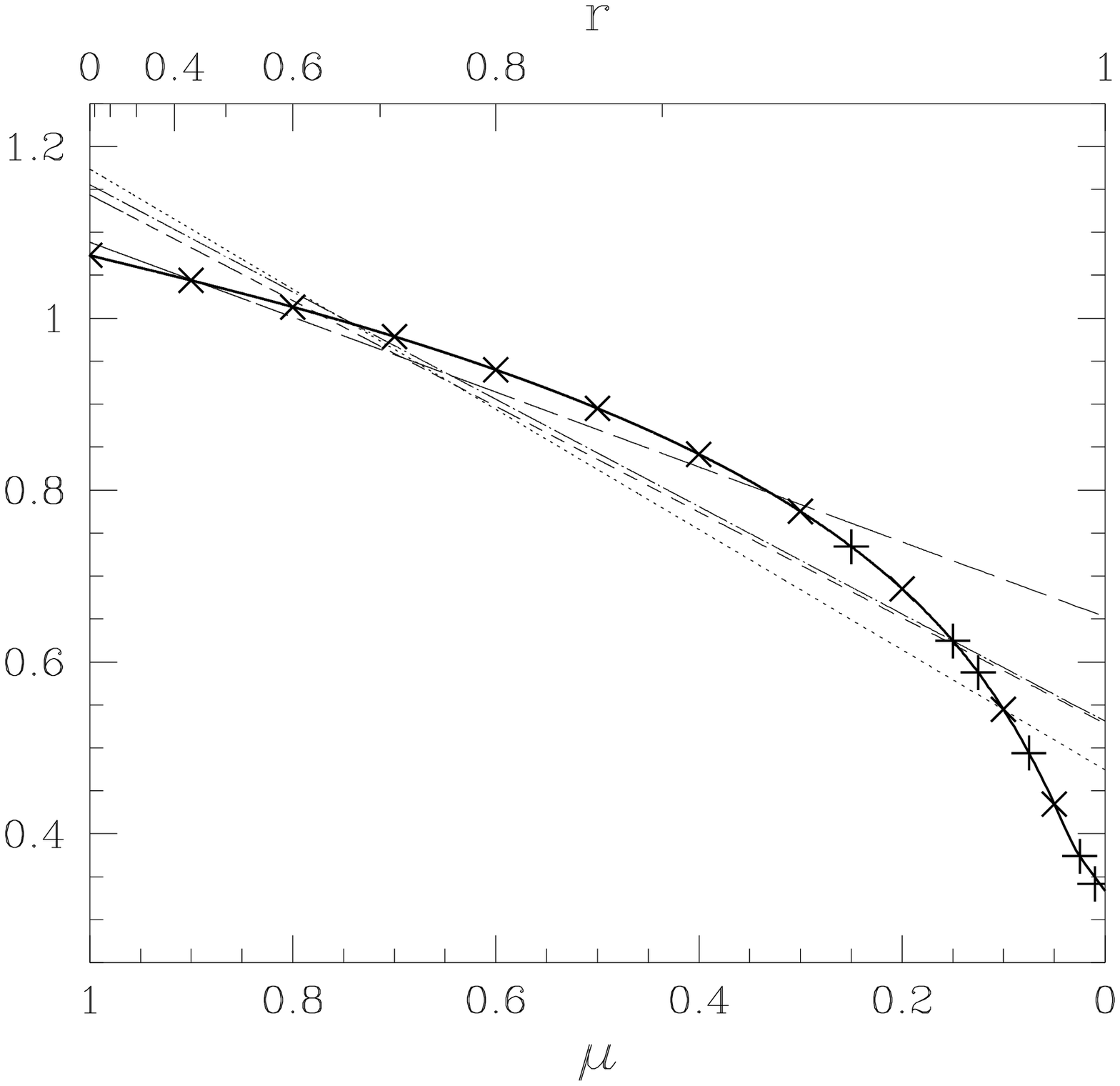}\\
\includegraphics[scale=.4]{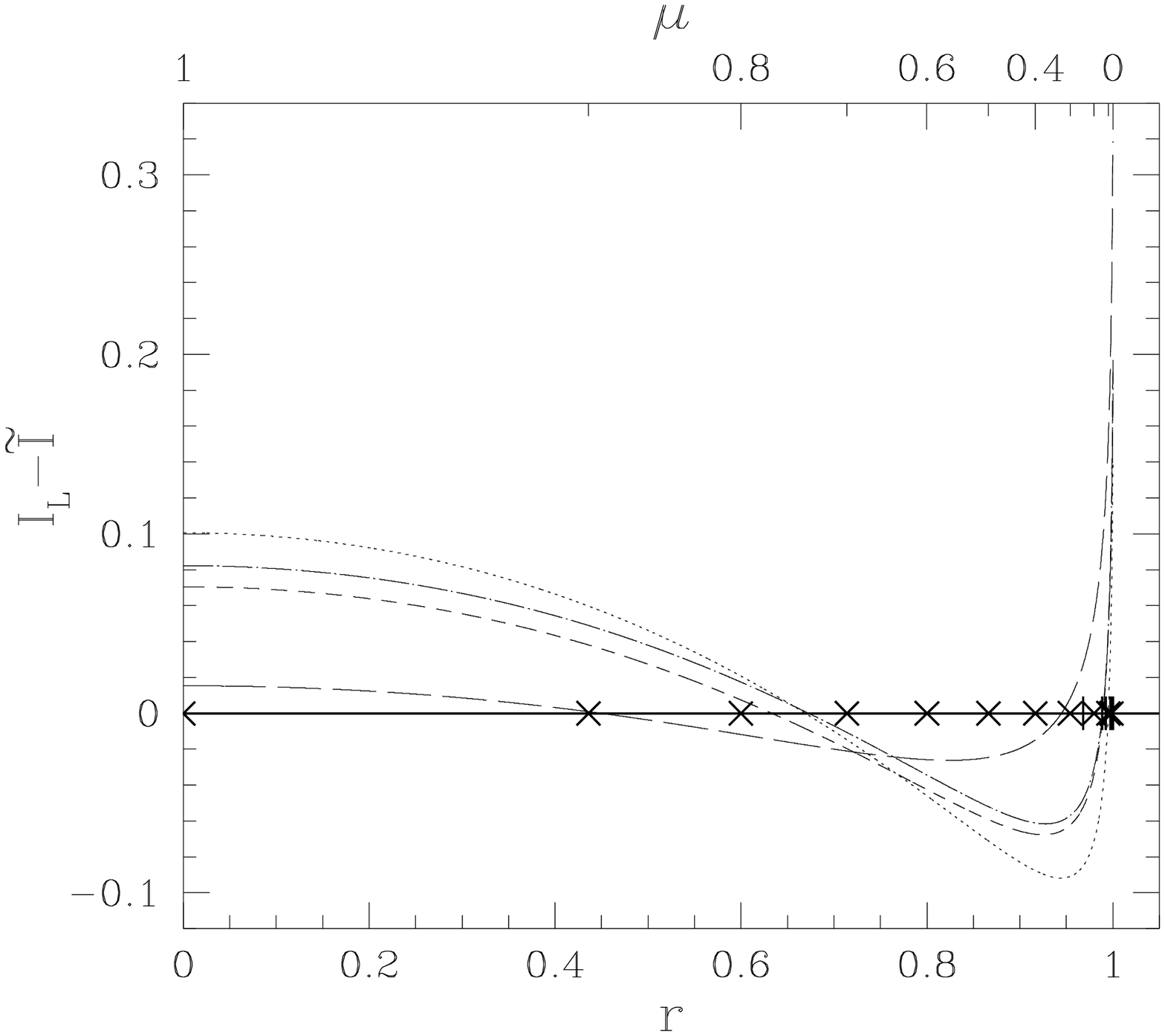}
\includegraphics[scale=.4]{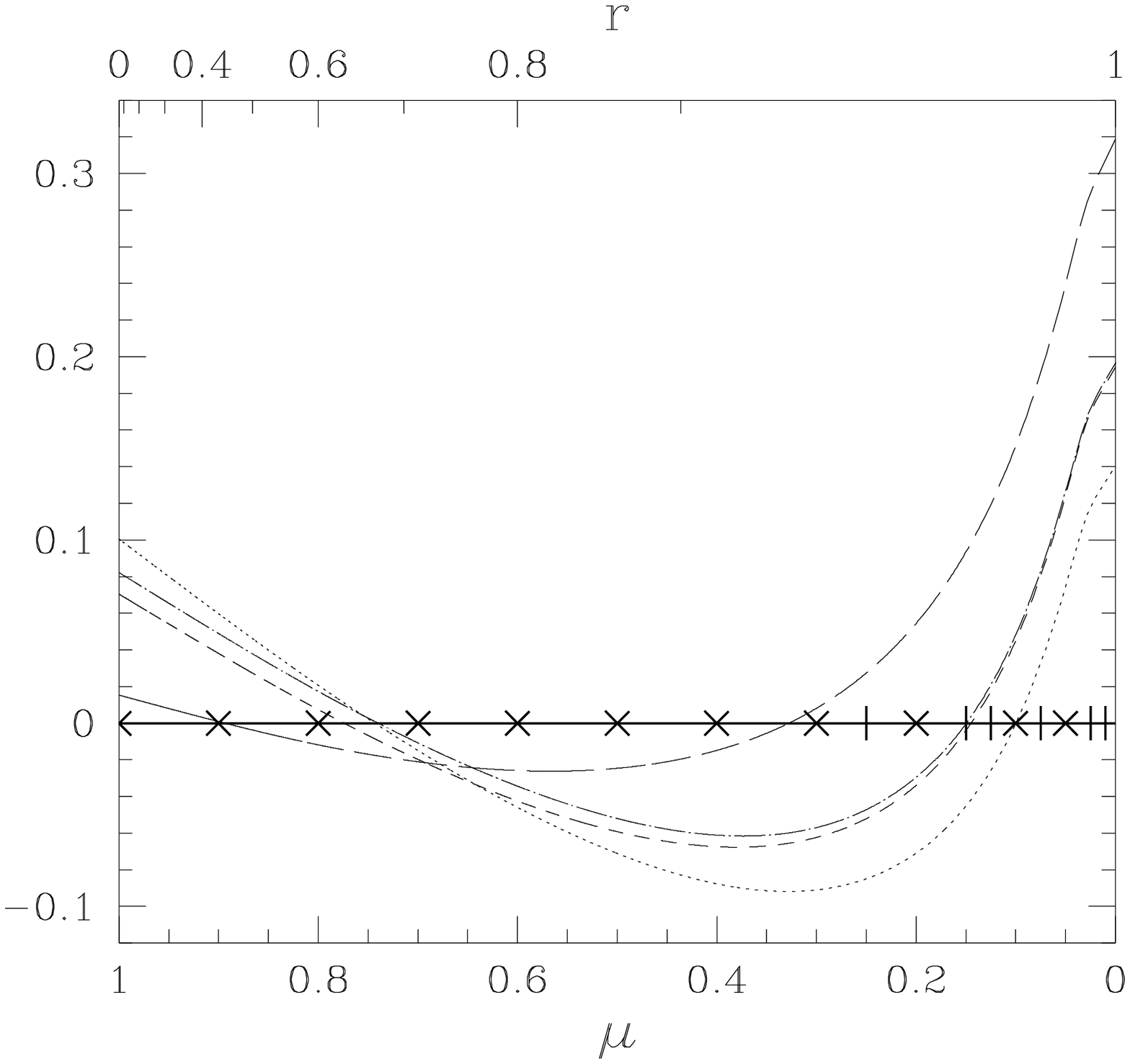}
\end{center}
\caption{Sample Kurucz ATLAS9 model intensity profile ($T_{eff}=4250\,K$, $log\,g=4.5$, $[Fe/H]=-3.5$, $V$ band) and its different linear limb darkening fits. Original 17 points marked by crosses (11 points) and pluses, lines include interpolated profile $\tilde{I}$ (bold solid), 17-point fit (dotted, linear coefficient $u_{17}=0.596$), 11-point fit (short-dashed, $u_{11}=0.538$), fit of interpolated $\tilde{I}$ by $\mu-$integration (dot-dashed, $u_{\mu}=0.540$), and fit of $\tilde{I}$ by $r-$integration (long-dashed, $u_r=0.401$). Intensity shown in top panels, absolute residuals of fits from $\tilde{I}$ in bottom panels. Left panels plotted as a function of $r$, right panels as a function of $\mu$, with respective alternate parameters marked along the upper axes.}
\label{fig:fits}
\enf

\clearpage
\bef[t]
\plotone{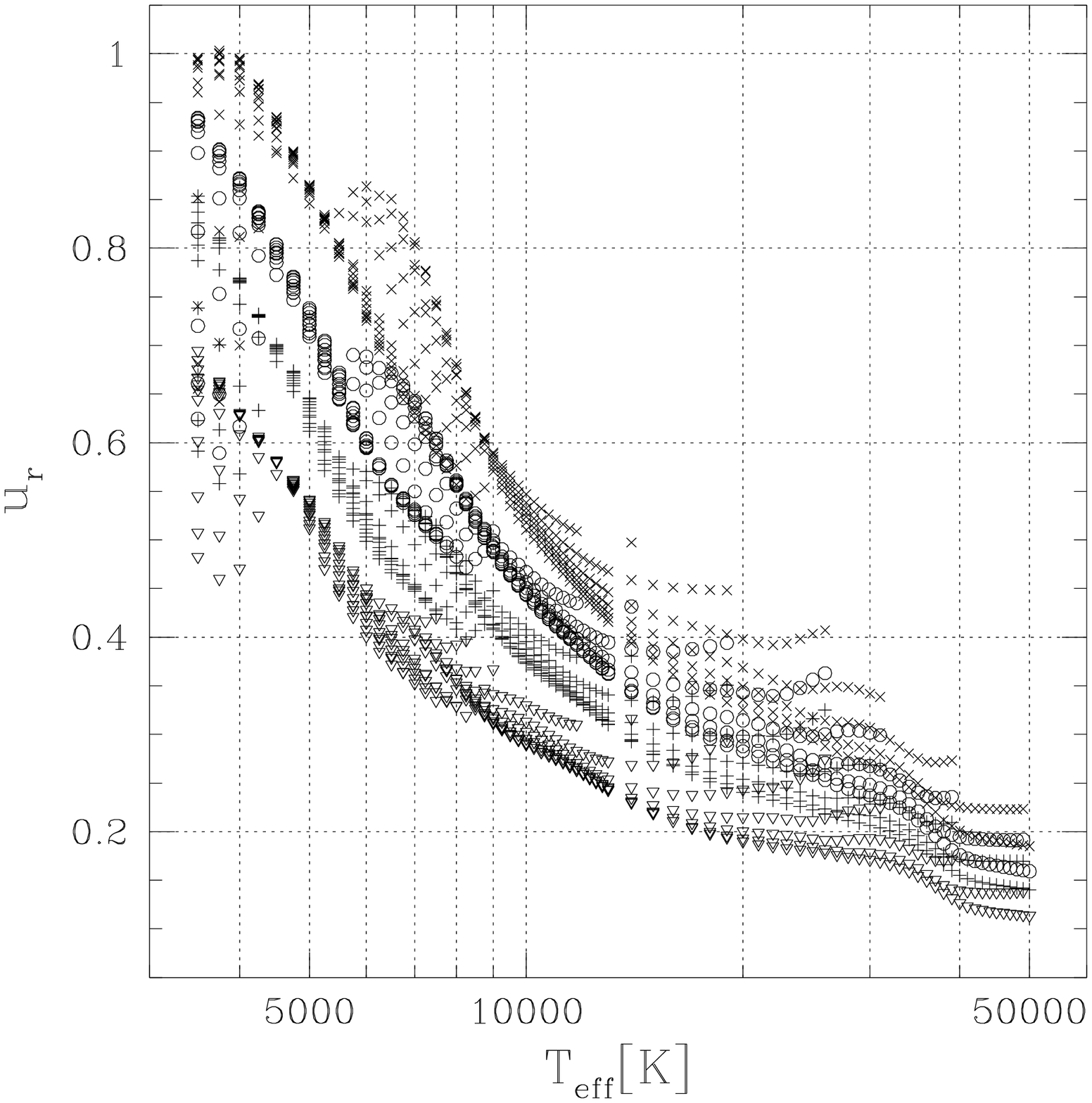}
\caption{Linear limb darkening coefficient $u_r$ of solar abundance, $v_t=2\,km\,s^{-1}$ Kurucz model atmospheres \citep{kur94} plotted as a function of effective temperature for bands $B$ (crosses), $V$ (circles), $R$ (pluses), $I$ (triangles).}
\label{fig:coeff-solarcomp}
\enf

\clearpage
\bef[t]
\begin{center}
\includegraphics[scale=.4]{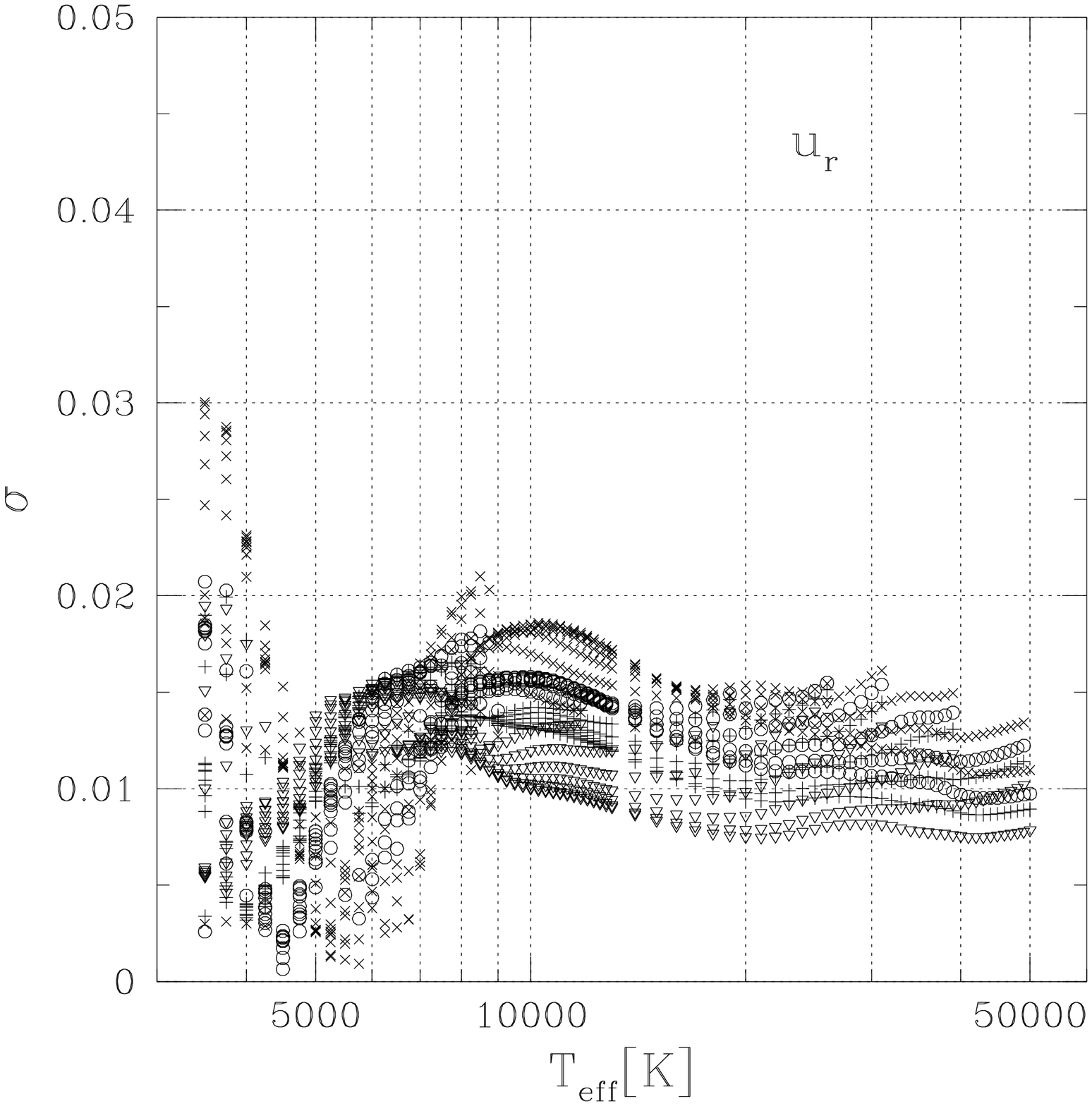}
\includegraphics[scale=.4]{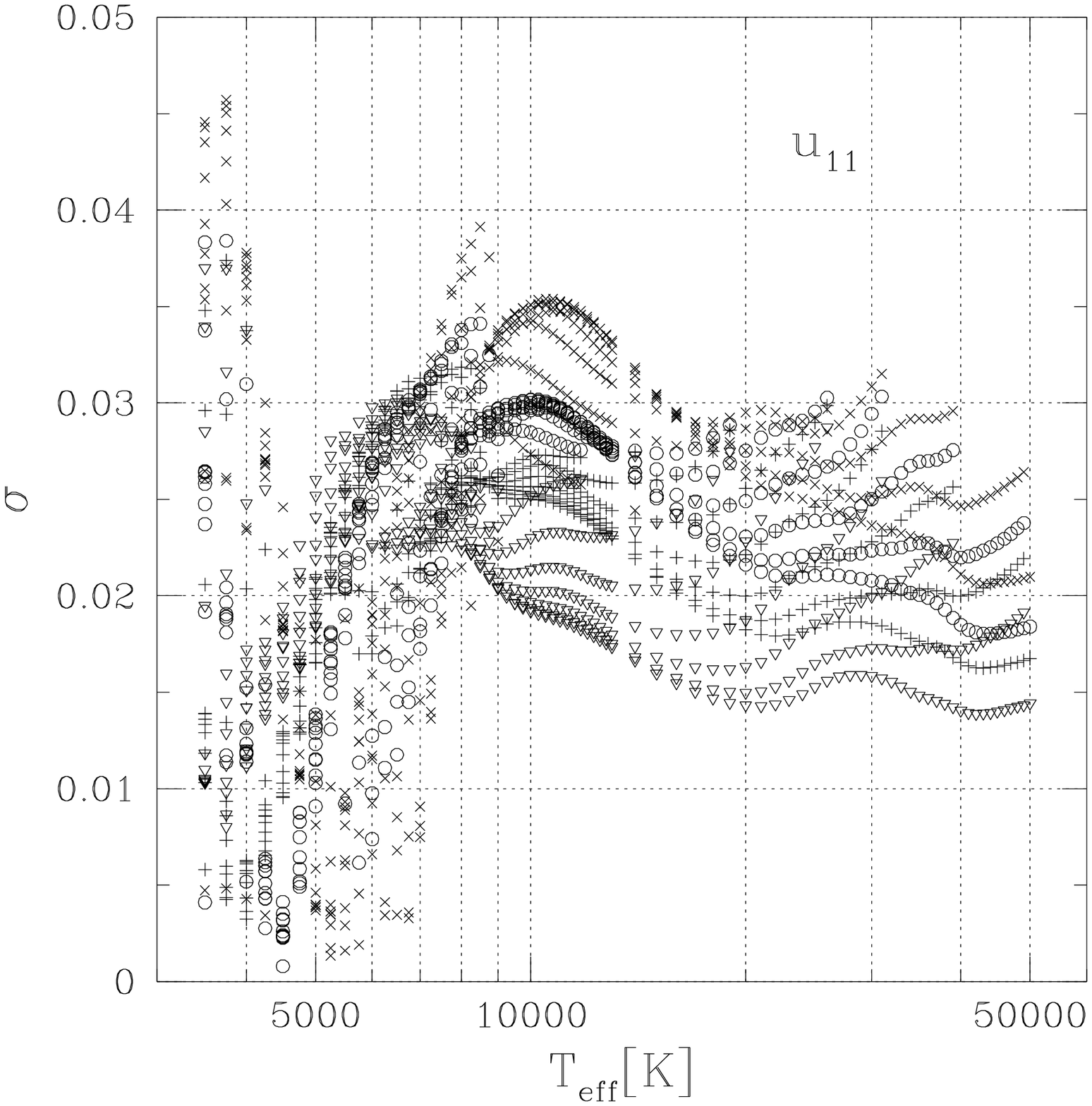}\\
\includegraphics[scale=.4]{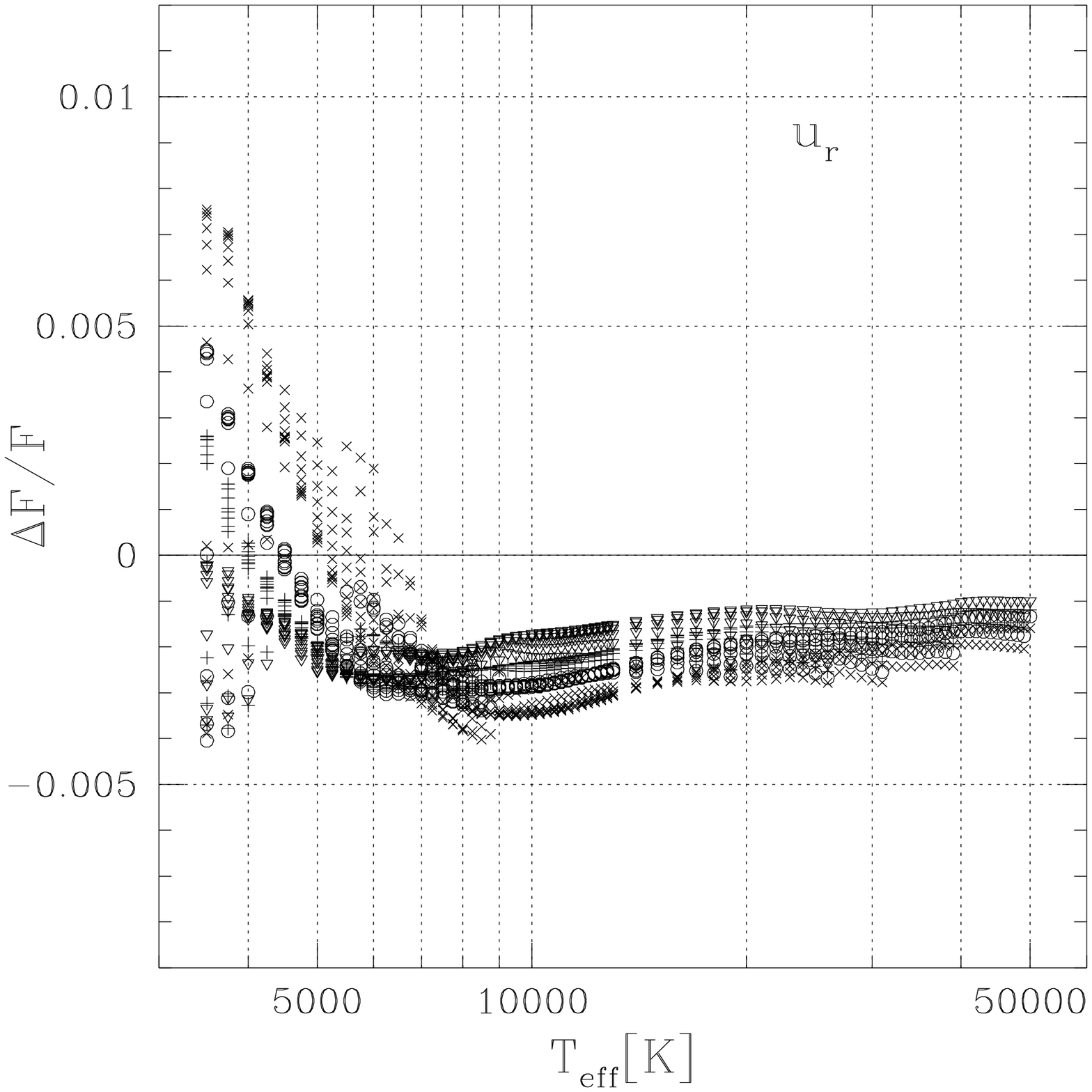}
\includegraphics[scale=.4]{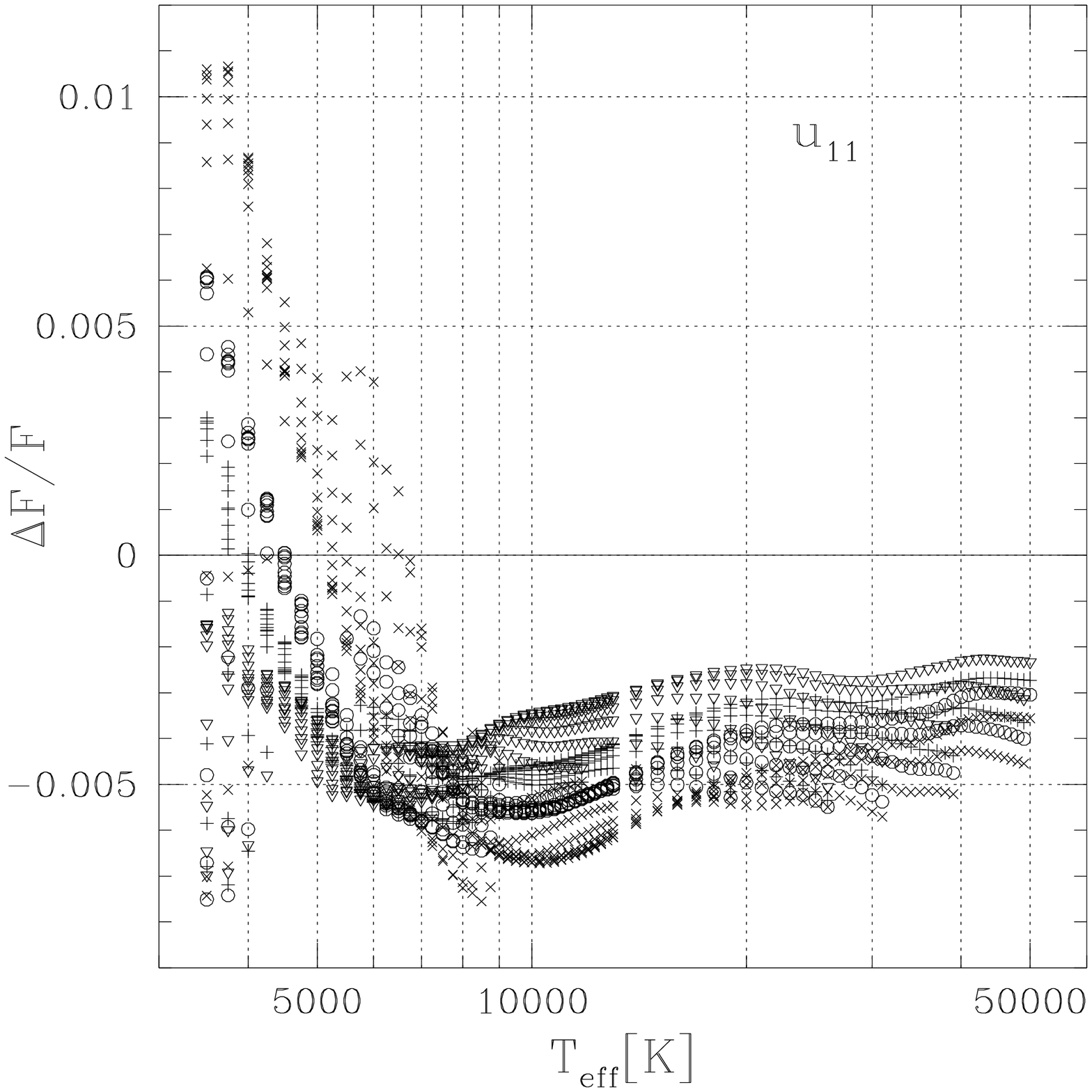}
\end{center}
\caption{Relative rms residuals $\sigma$ and relative flux excesses $\Delta F/F$ for the $r-$integrated limb darkening fits from Figure~\ref{fig:coeff-solarcomp} (left column) and for 11-point fits (right column) plotted as a function of effective temperature. Range of model atmospheres and symbols for photometric bands as in Figure~\ref{fig:coeff-solarcomp}.}
\label{fig:resid-solarcomp}
\enf

\clearpage
\bef[t]
\plotone{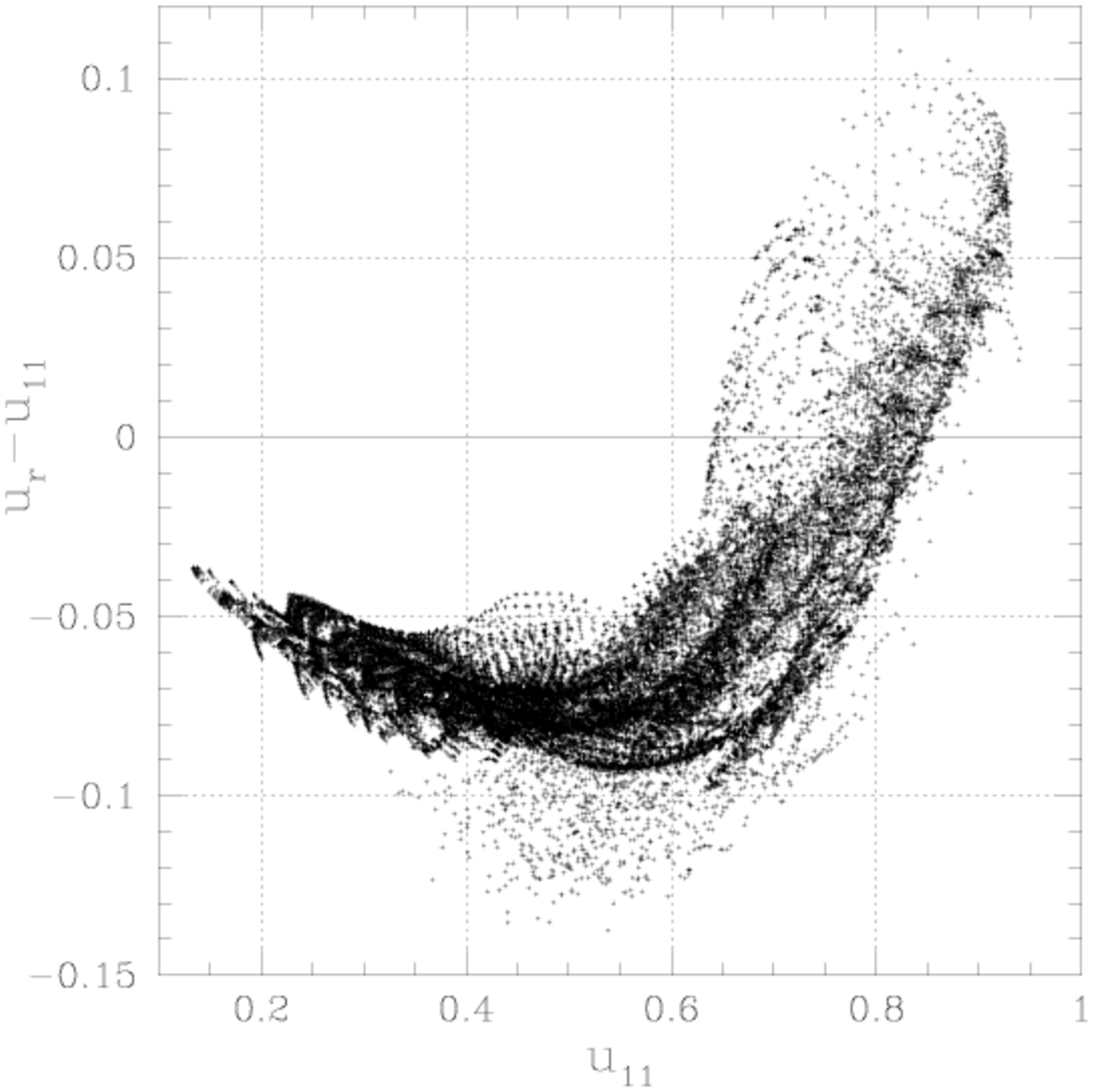}
\caption{Difference between linear limb darkening coefficients $u_r$ and $u_{11}$ plotted as a function of $u_{11}$ for full range of models (see text for details).}
\label{fig:allcoeffdiff}
\enf

\clearpage
\bef[t]
\begin{center}
\includegraphics[scale=.4]{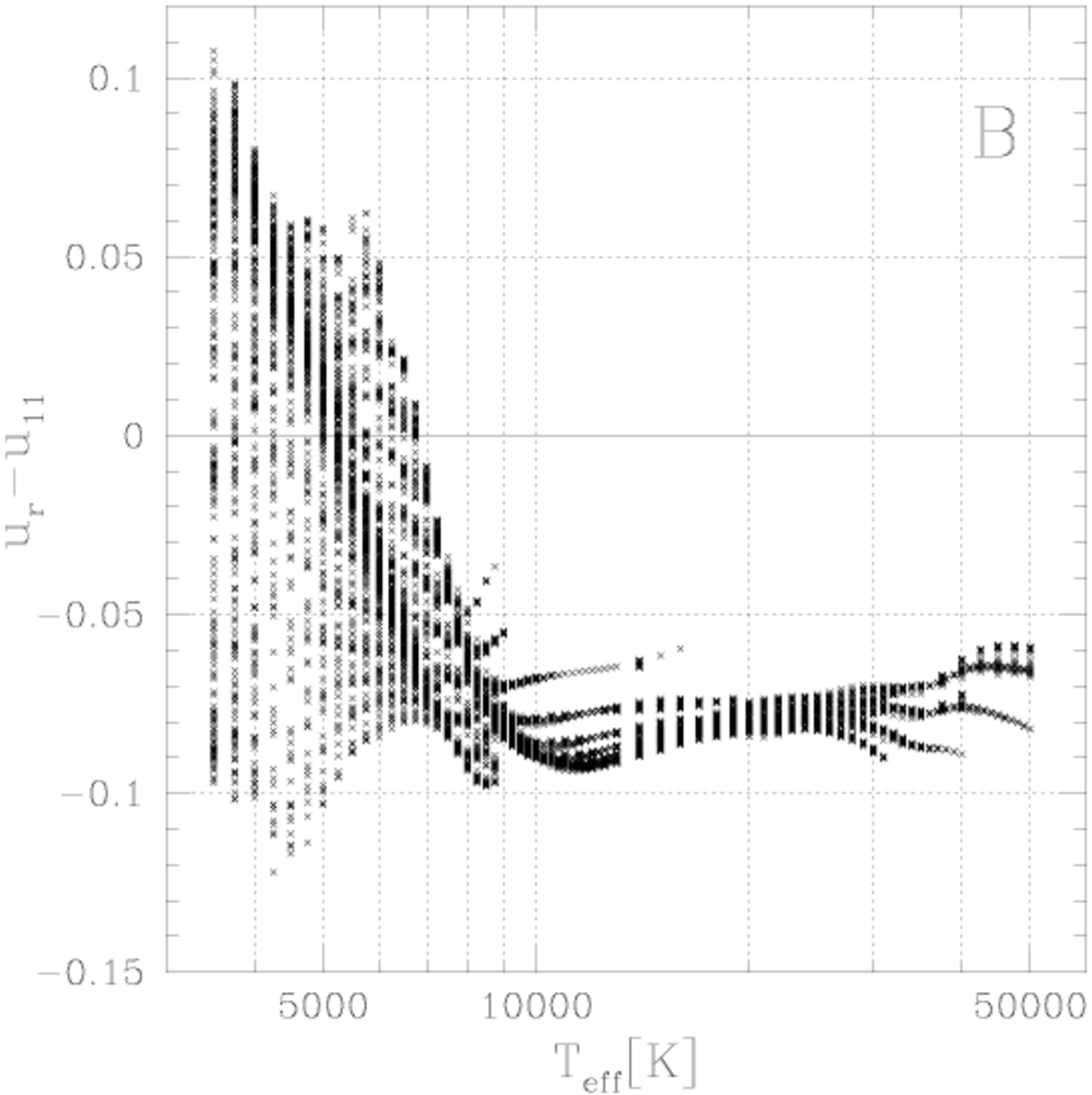}
\includegraphics[scale=.4]{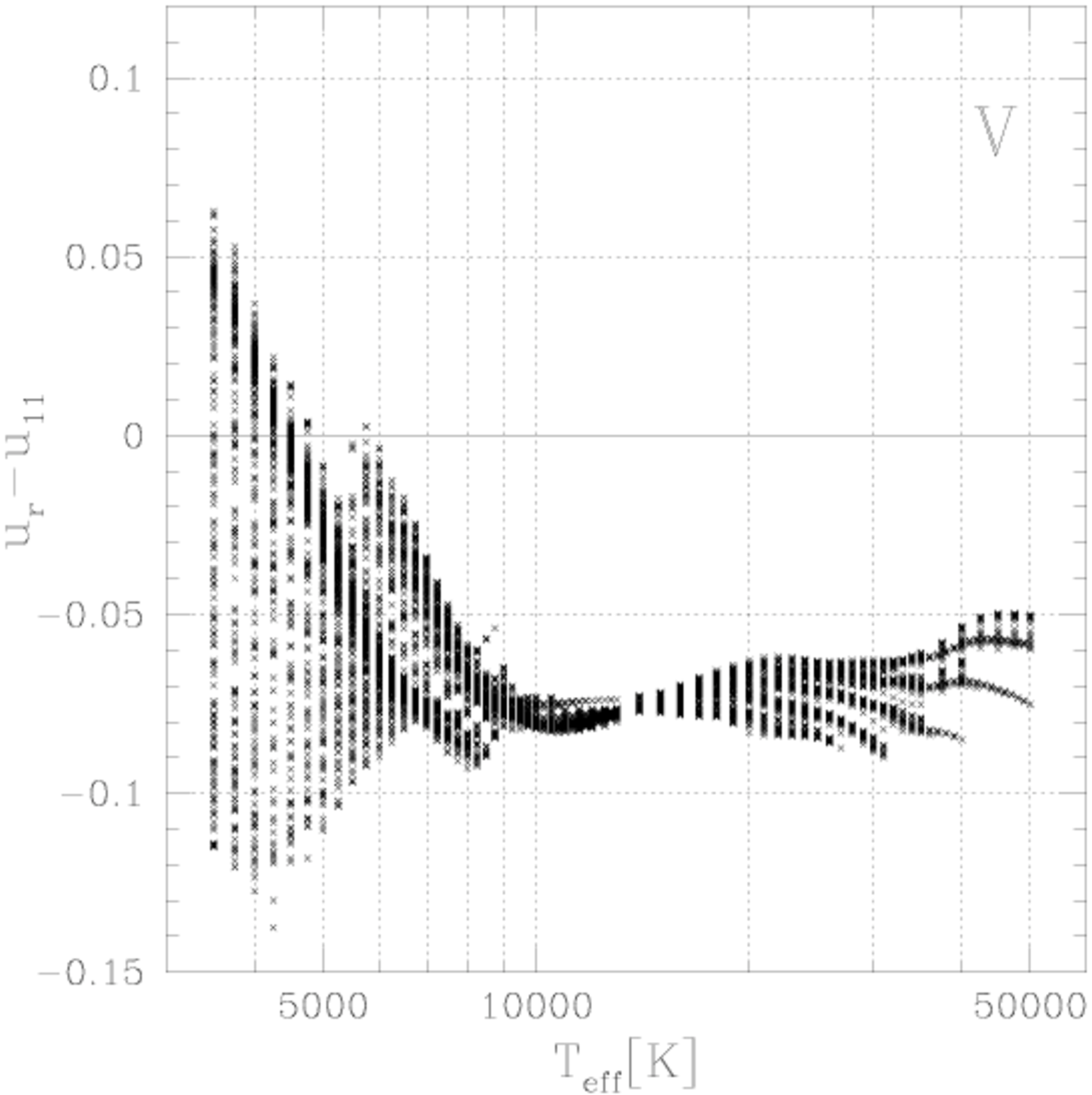}\\
\includegraphics[scale=.4]{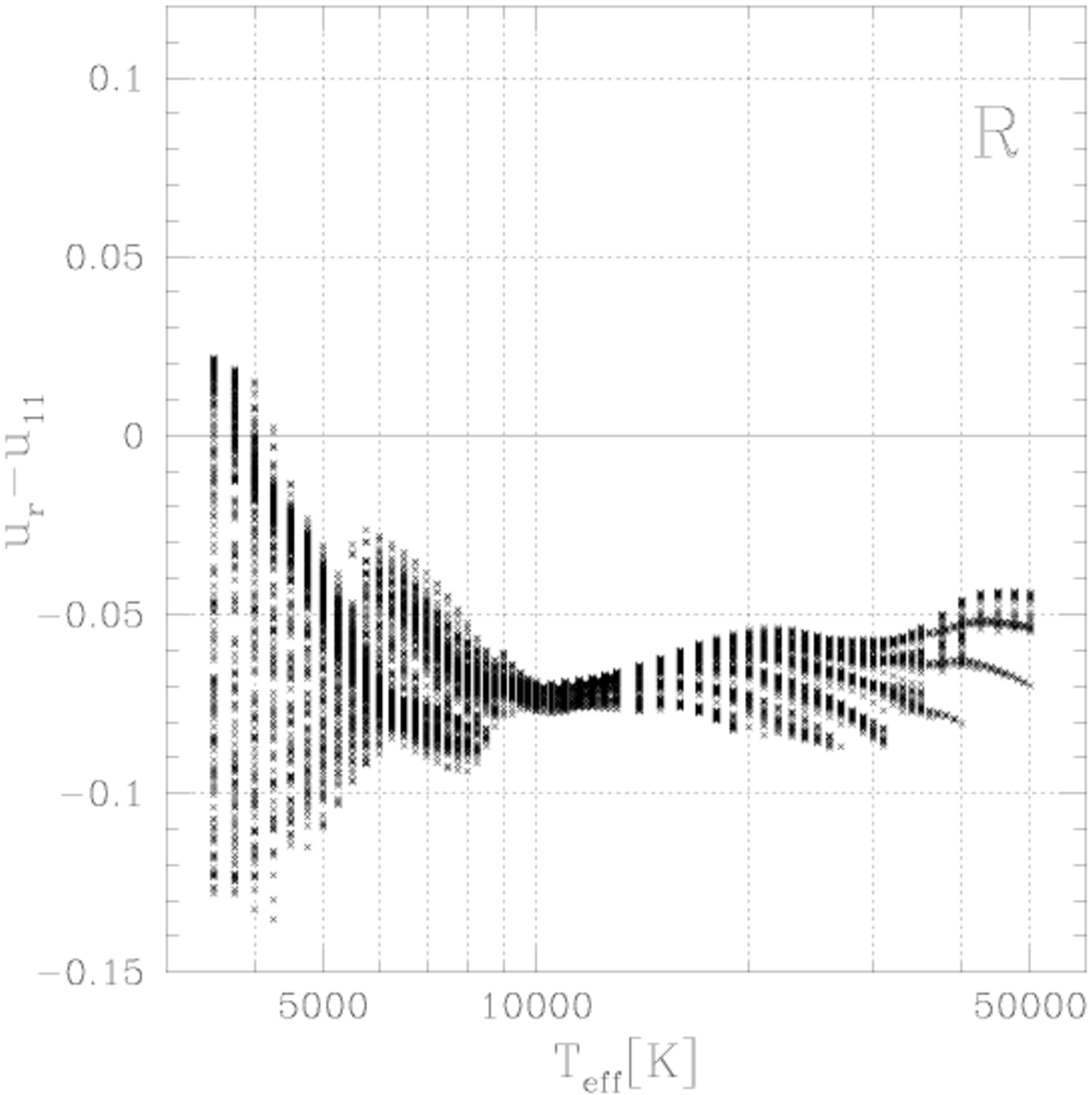}
\includegraphics[scale=.4]{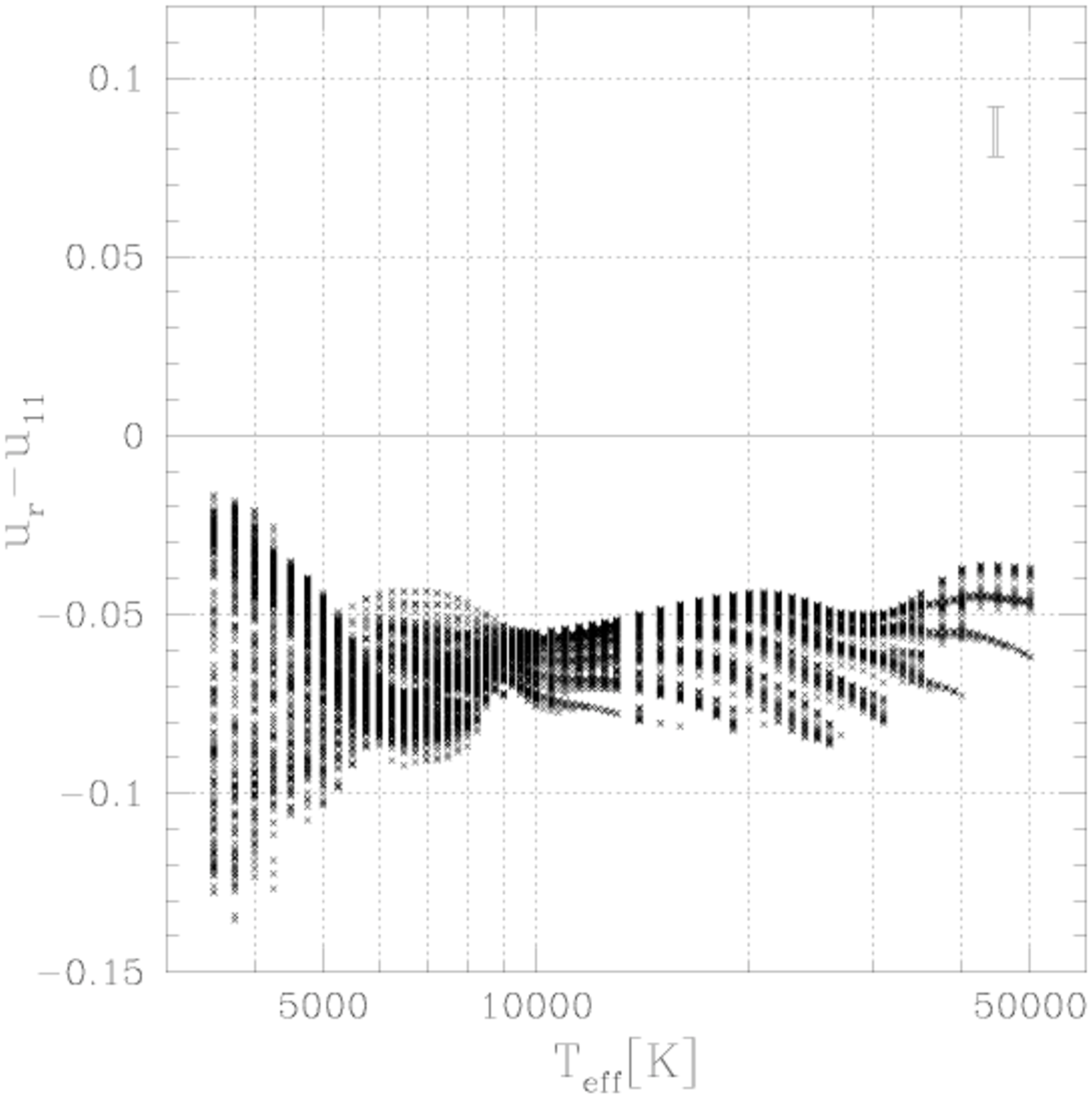}
\end{center}
\caption{Difference between linear limb darkening coefficients $u_r$ and $u_{11}$ for full range of models plotted as a function of $T_{eff}$ separately for bands $B$ (top left), $V$ (top right), $R$ (bottom left), $I$ (bottom right).}
\label{fig:bandcoeffdiff}
\enf

\clearpage
\bef[t]
\begin{center}
\includegraphics[scale=.4]{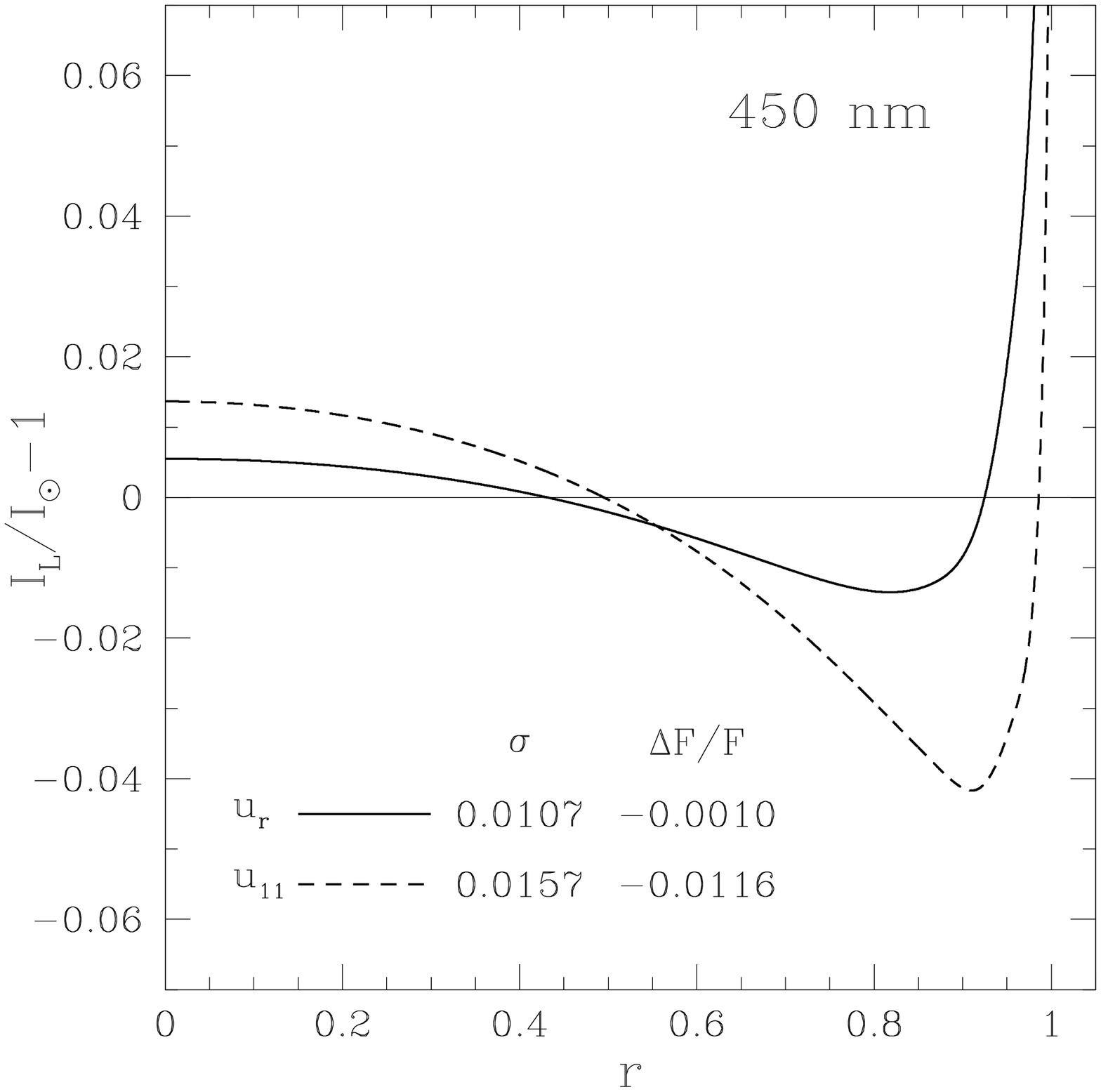}
\includegraphics[scale=.4]{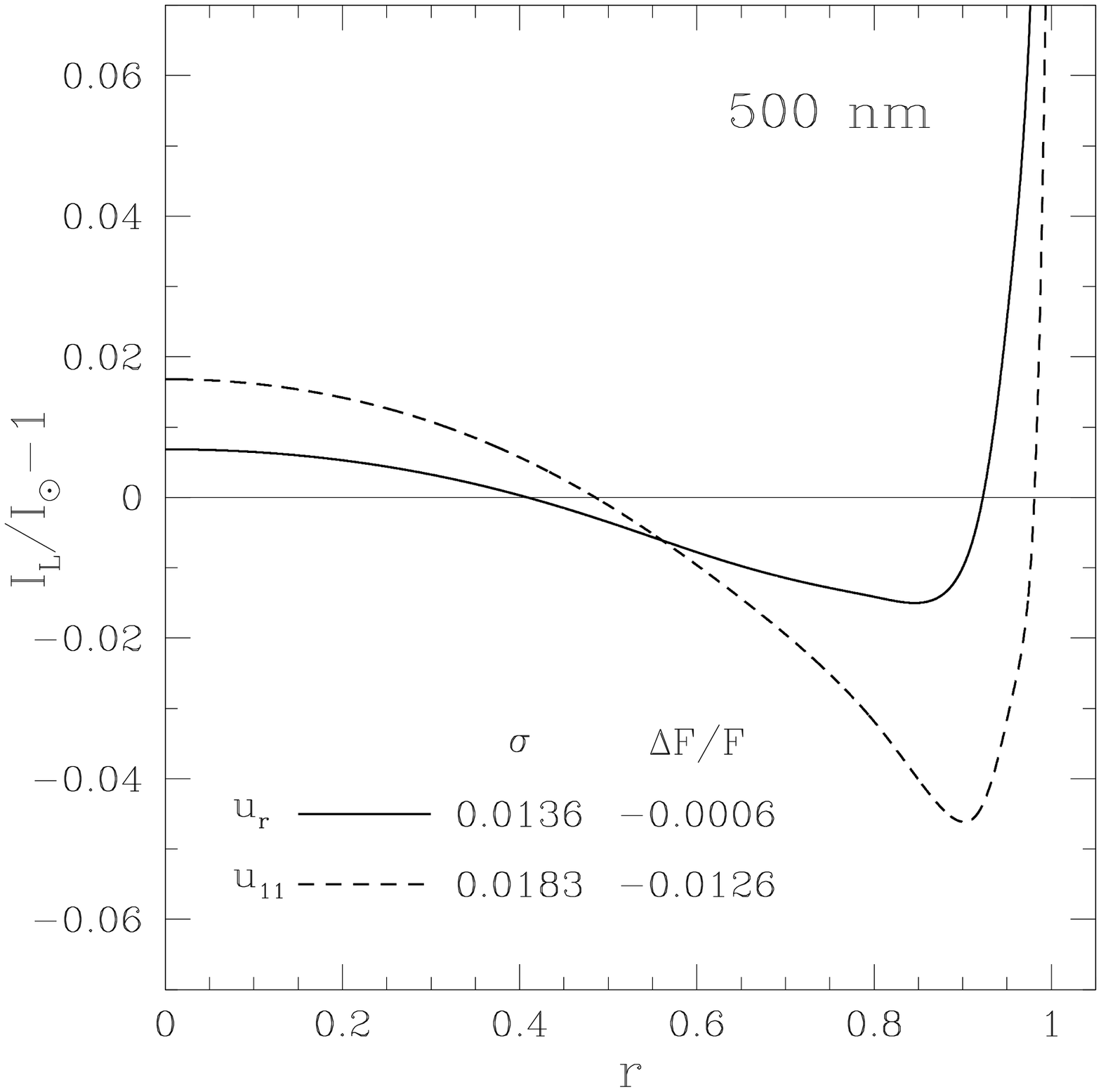}\\
\includegraphics[scale=.4]{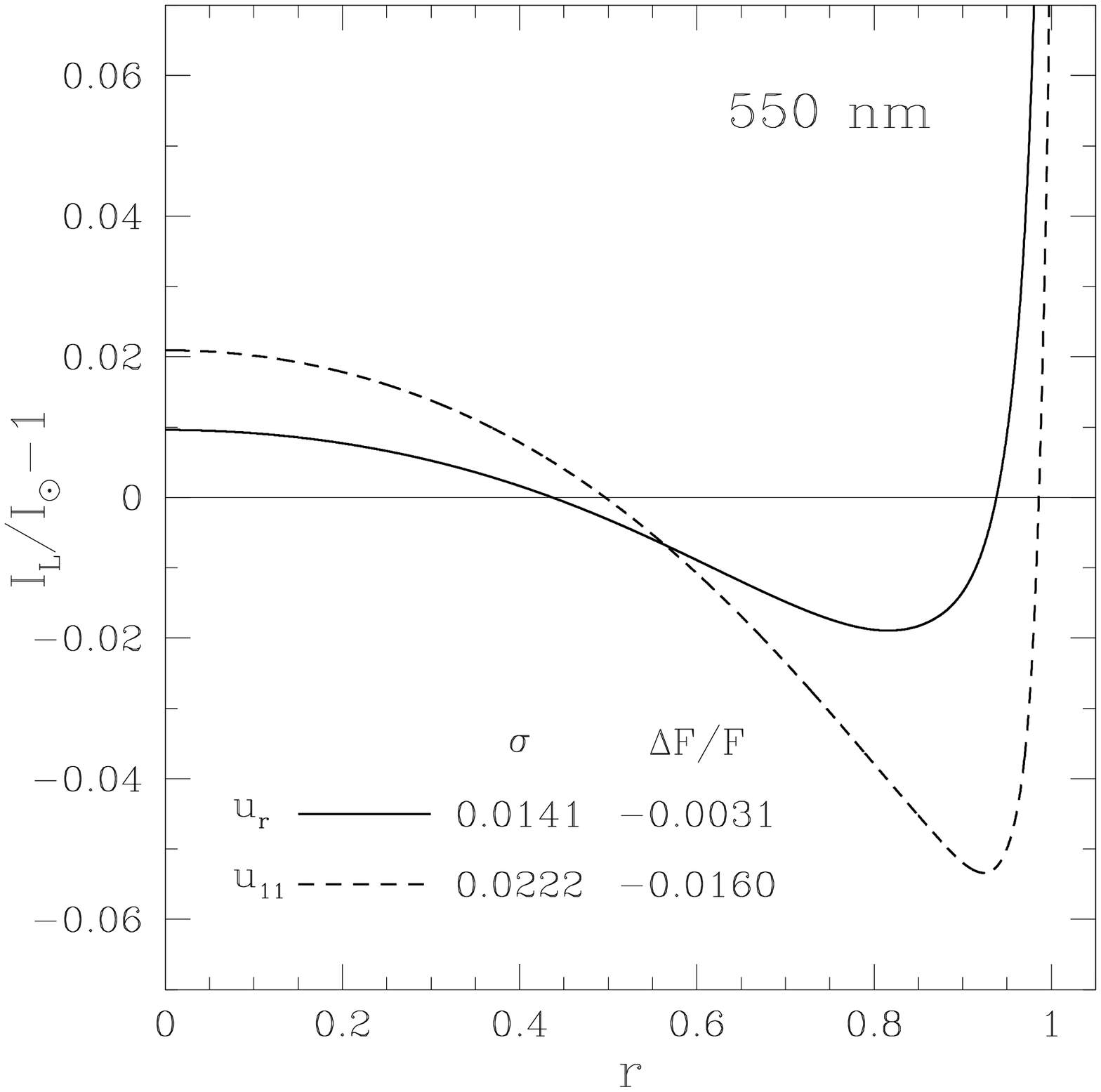}
\includegraphics[scale=.4]{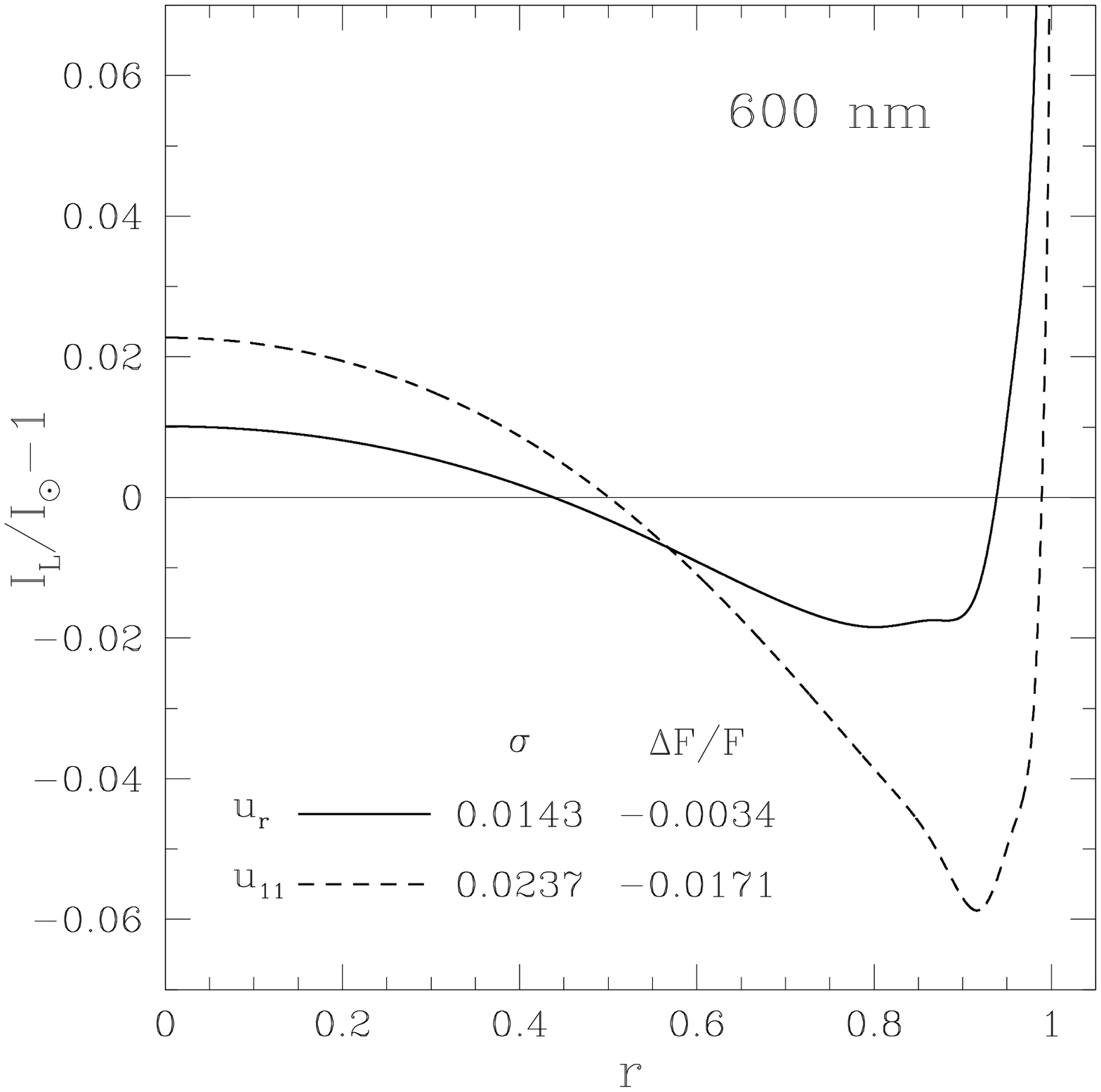}
\end{center}
\caption{Relative residual $I_L/I_\odot-1$ of linear limb darkening law $I_L$ with theoretical coefficients $u_r$ (solid line) and $u_{11}$ (dashed line) compared with solar limb darkening $I_\odot$ \citep{cox00} for four wavelengths (marked in upper right corners of the panels), plotted as a function of radial position $r$. Relative rms residuals $\sigma$ and relative flux excesses $\Delta F/F$ for both fits are given in each panel.}
\label{fig:sun}
\enf

\end{document}